\documentclass[a4paper,fleqn,review]{cas-dc}
\usepackage{amsmath,amssymb,amsfonts}
\usepackage{graphicx}
\usepackage{soul}
\usepackage{natbib}
\usepackage[most]{tcolorbox}
\usepackage{makecell} 
\setcitestyle{authoryear,round}
\usepackage{enumitem}
\usepackage{diagbox}
\usepackage{listings}
\usepackage{supertabular}
\usepackage{lipsum} 
\usepackage{hyperref}
\usepackage{booktabs} 
\usepackage{pifont}

\usepackage{textcomp}
\usepackage{xcolor,colortbl}
\usepackage{caption}
\usepackage{float}
\usepackage{subcaption}
\RequirePackage{tikz}
\usepackage{xfrac}
\usepackage{multirow}
\usepackage{multicol}
\usepackage{balance}
\usepackage{enumitem}
\usepackage{algorithm}
\usepackage{algorithmic}
\usepackage[underline=false]{pgf-umlsd}
\usepackage{mathtools}
\begin{document}
\shorttitle{What You See Is Not What You Execute Memory-Based Runtime SBOM Generation for Supply Chain Security}
\shortauthors{Hala Ali, Andrew Case,  Irfan Ahmed}  
\title[mode = title]{What You See Is Not What You Execute: Memory-Based Runtime SBOM Generation for Supply Chain Security} 

\author[1]{Hala Ali}
\cormark[1]
\ead{alih16@vcu.edu}
\affiliation[1]{organization={Department of Computer Science},
                addressline={Virginia Commonwealth University}, 
                country={USA}}

\author[2]{Andrew Case}
\ead{andrew@dfir.org}
\affiliation[2]{organization={Volatility Foundation},
                country={USA}}
\author[1]{Irfan Ahmed}
\ead{iahmed3@vcu.edu}

\cortext[cor1]{Corresponding author}

\begin{abstract}
Modern software development relies heavily on third-party components from public repositories, expanding the software supply chain attack surface. In response to these growing risks,  federal initiatives have advanced the Software Bill of Materials (SBOM) as a standardized mechanism for improving transparency by describing software components, dependencies, and their relationships. However, SBOMs built from metadata or filesystem artifacts fail to capture the components loaded and executed at runtime, especially in dynamic ecosystems such as Python. 
Moreover, generating runtime SBOMs through instrumentation requires monitoring to be deployed in advance and the system to remain observable throughout execution. Such conditions are  difficult to satisfy in production environments and incident-response scenarios. Volatile memory, in contrast, provides a reliable source for recovering the actual runtime state of a running application without requiring prior instrumentation.
Therefore, this paper presents \textsc{MEM-SBOM}, the first memory forensics framework that generates SBOMs directly from the runtime state of Python applications. It recovers the modules from the interpreter’s internal structures, resolves package versions, and analyzes bytecode to build dependency graphs and identify reachable vulnerable functions. We implemented \textsc{MEM-SBOM} as a suite of Volatility 3 plugins and evaluated it against  51 real-world Python applications. It achieves 100\% extraction accuracy, identifies \textit{Streamlit} as the only application that calls the vulnerable routines of the \textit{tornado} dependency, and recovers all runtime packages missed by existing SBOM tools, providing more accurate dependency graphs and better vulnerability assessment. These capabilities make \textsc{MEM-SBOM} a practical foundation for software supply chain security and incident response by providing a forensically sound runtime view of what is executed on a system.
\end{abstract}
\maketitle

\section{Introduction}
\label{s1}
The widespread adoption of third-party libraries and frameworks from public repositories has become integral to modern software development, enabling rapid feature integration while simultaneously expanding the software supply chain attack surface \citep{nahum2024ossintegrity,cisa2025sbom_3, ncsc}. Adversaries are increasingly exploiting these repositories to distribute malicious packages that compromise downstream applications \citep{vu2023bad}. In 2024, more than 700,000 malicious packages were discovered in public registries, a 156\% increase over the previous year \citep{sonatype_supply_chain}. The official Python repository, \textit{PyPI}, has repeatedly faced such incidents, with more than 12,000 malicious packages removed in 2022 alone \citep{pypi_inbound_malware_2023}, yet malicious uploads persist despite repository-level scanning \citep{samaana2025machine, kampourakis2025cracks}. Examples include \texttt{zlibxjson v8.2}, which embedded credential-stealing scripts that exfiltrated cookies and authentication tokens \citep{fortiguard2024zlibxjson}, and \texttt{fshec2}, which concealed credential theft and remote command handling within compiled bytecode (\texttt{.pyc}) to evade detection \citep{fshec2}. In April 2025, additional malicious packages (\texttt{bitcoinlibdbfix}, \texttt{bitcoinlib-dev}, and \texttt{disgrasya}) were downloaded more than 39,000 times before being removed, with the latter embedding an automated script for credit-card fraud \citep{lakshmanan2025pypi}. Even trusted frameworks can introduce critical vulnerabilities, such as the recent SQL-injection flaw in \textit{Django} (CVE-2025-64459), with affected versions still publicly available on PyPI \citep{django_cve2025_64459} at the time of writing.

These recurring incidents have highlighted the urgent need for software supply chain transparency, prompting regulatory and industry efforts to improve visibility into software composition. In response, initiatives led by the Cybersecurity and Infrastructure Security Agency (CISA) and the National Telecommunications and Information Administration (NTIA) have advanced the Software Bill of Materials (SBOM) as a standardized mechanism for software transparency and risk management \citep{cisa_sbom_practices}. An SBOM is a structured inventory of software components and dependencies, specifying their metadata (e.g., name, version, and license) and relationships \citep{cisa2024sbom_2}. Based on CISA’s guidance, SBOMs have become an increasingly expected requirement of federal software procurement \citep{cisa2025sbom_1}. As a result, several tools have emerged to automate SBOM generation across ecosystems. However, these tools remain constrained by their reliance on static metadata, such as build manifests and package files (e.g., \texttt{requirements.txt}, \texttt{pyproject.toml}, and \texttt{setup.py}), which describe what was built rather than what actually executes.  Prior studies show that these metadata-driven tools generate incomplete and inaccurate SBOMs due to inconsistent format support, parser divergence, and weak handling of dynamic or environment-specific dependencies \citep{yu2024correctness, dalia2024sbom, cofano2024sbom, xia2023empirical, stalnaker2024boms}.  Furthermore, metadata files themselves can be compromised, as attackers may modify dependency specifications to mislead SBOM tools into producing incorrect component and version information \citep{zahan2023software}.

Recent efforts have extended SBOM generation to the software deployment phase, including container-level analysis \citep{kawaguchi2024deployment}, package-manager integration \citep{benedetti2024impact}, and integrity verification of deployed components \citep{sharma2024sbom}. Although these approaches improve accuracy compared to static build-phase methods, they still rely on filesystem artifacts and package managers, which reflect the software's state at install. This limitation becomes even more evident in dynamic, interpreted languages where dependencies are resolved during execution rather than at build time.
Python presents a particularly challenging and representative example of such dynamic ecosystems due to its highly flexible import mechanism, which allows modules to be loaded at runtime through conditional, delayed, or reflective imports \citep{ali2025memory}. This flexibility causes different frameworks and applications to load and retain dependencies in distinct ways. For example, when upgrading shared dependencies such as \textit{asgiref}, \textit{Django}’s \textit{StatReloader} dynamically loads the newer version into memory during execution, whereas \textit{Celery} lacks an auto-reload mechanism and continues using the older version already loaded at startup.


While runtime SBOMs aim to capture the active components of running applications through system instrumentation \citep{mirakhorli2024landscape}, this approach assumes the system can be safely monitored throughout execution. In production environments, such monitoring introduces operational overhead and risks perturbing the runtime state under investigation \cite{beyer2016site, reichelt2024overhead}. Moreover, in incident-response scenarios, compromised systems cannot be safely paused, debugged, or retroactively instrumented. Volatile memory, in contrast, provides a reliable source for recovering the actual runtime components and dependencies, and thus for generating accurate runtime SBOMs.

Therefore, in this paper, we propose \textsc{MEM-SBOM}, the first memory forensics framework that generates SBOMs directly from the runtime state of Python applications. It traverses the interpreter’s internal structures, including module registries, thread contexts, the garbage collector, arenas, and heap regions, to extract all modules resident in memory. This multi-layered architecture ensures completeness and forensic soundness even in the presence of evasion techniques, such as module deletion and overwrite, import-hook interception, import-bypassing module creation, sub-interpreter injection, and garbage collector manipulation. Modules recovered from all layers and application processes are then aggregated, filtered to exclude built-in and standard-library components, and grouped by their top-level packages.  To determine the versions of the extracted third-party packages, \textsc{MEM-SBOM} inspects version attributes across all modules, correlates them with installation metadata, and queries PyPI when necessary before serializing these packages into a CycloneDX-compliant SBOM. Beyond generating SBOMs, \textsc{MEM-SBOM} analyzes the Python objects associated with each package and disassembles the bytecode of its functions to construct dependency graphs that capture both direct and transitive runtime relationships. Leveraging the resulting SBOMs and graphs and integrating public vulnerability data, the framework performs fine-grained, function-level reachability analysis to distinguish the exploitable code paths at runtime.

We implemented \textsc{MEM-SBOM} as a suite of new Volatility 3 plugins  \citep{volatility_framework}. We then evaluated it across 51 real-world Python applications spanning 23 diverse categories, including CLI tools, web frameworks, machine learning platforms, workflow schedulers, and visualization tools. Validated against a dual static and runtime ground truth, \textsc{MEM-SBOM} accurately extracts the metadata cached for installed packages and recovers all application packages, including those dynamically loaded at runtime, across heterogeneous execution environments. It further identifies ten cases where the version embedded in the loaded code differs from the version declared at installation time. In a \textit{tornado} case study, function-level reachability analysis shows that only \textit{Streamlit}, among six \textit{tornado}-dependent applications, invokes the vulnerable routines at runtime, eliminating 83.3\% of false-positive vulnerability reports. Finally, our comparison against eight state-of-the-art SBOM tools shows that none of them support all Python metadata configurations or capture dynamically loaded packages, resulting in incomplete SBOMs, partially constructed dependency graphs, and ultimately inaccurate vulnerability reports.

Our contributions are as follows:
\begin{itemize}
    \item We propose \textsc{MEM-SBOM}, the first framework to generate SBOMs directly from the runtime state of Python applications, providing an execution-grounded view of software components.

    \item We characterize evasion techniques that conceal Python modules during execution and design a robust multi-layered extraction architecture capable of detecting and recovering all modules.

    \item We construct runtime dependency graphs and perform function-level vulnerability reachability analysis, identifying exploitable code paths.
    
    \item We evaluate \textsc{MEM-SBOM} across 51 real-world Python applications, achieving higher accuracy and vulnerability precision than existing SBOM tools.

\end{itemize}

The remainder of this paper is organized as follows. Section \ref{s2} provides the background, and Section \ref{s3} summarizes related work. Section \ref{s4} outlines the key challenges, while Section \ref{s5} explains the methodology. Section \ref{s6} presents the evaluation, Section \ref{s7} discusses limitations and future directions, and Section \ref{s8} concludes the paper.

\section{Background}
\label{s2}
This section presents SBOM types and standards, followed by an overview of the import mechanism and memory analysis of the Python runtime.
\subsection{SBOM Fundamentals} 
\label{s2_1}
SBOMs can be generated at different stages of the software lifecycle, each providing a distinct level of visibility into software composition \citep{cisa_sbom_types}.  \textbf{Build SBOMs} describe the components, dependencies, and build metadata incorporated during compilation or packaging. \textbf{Deployed SBOMs} enumerate the components installed and configured within operational environments. \textbf{Runtime SBOMs}, in turn, capture the components actively loaded during execution, revealing dynamic and environment-specific dependencies that are invisible to static analysis. SBOMs are commonly represented using two standards. \textbf{SPDX}, maintained by the Linux Foundation, focuses on license compliance, copyright attribution, and file-level provenance \citep{spdx}. \textbf{CycloneDX}, developed by OWASP, is a lightweight, security-oriented standard that emphasizes dependency relationships, component provenance, and vulnerability correlation \citep{cyclonedx}.

\subsection{Import Mechanism of Python}
\label{s2_2}
The import mechanism of Python follows a multi-stage resolution process formalized in \textit{PEP 451} and \textit{PEP 420} \citep{pep451, pep420}. When the interpreter encounters an \texttt{import}  statement, it first checks its per-interpreter registry (\texttt{sys.modules}), which maps module names to loaded \texttt{PyModuleObject} instances.  If the module is not found, the import machinery proceeds through finder objects listed in \texttt{sys.meta\_path}. Each finder implements \texttt{find\_spec()} to locate the requested module and, upon success, returns a \texttt{ModuleSpec} that encapsulates the module’s name, origin, loader, and other import metadata.  Using this specification, the interpreter calls \texttt{loader.create\_module(spec)}  if provided, or creates a new module object otherwise, registers it in \texttt{sys.modules}, and executes it via \texttt{loader.exec\_module(module)} to populate its namespace.  Built-in and frozen modules use non-filesystem origins (e.g., "built-in" or "frozen"), whereas file-based modules populate attributes, such as \texttt{\_\_file\_\_} and \texttt{\_\_cached\_\_}. This spec-driven model replaces the legacy \texttt{load\_module()} API,  separating module discovery (\texttt{find\_spec()}) from execution (\texttt{exec\_module()}) and exposing import state through the \texttt{\_\_spec\_\_} attribute for reliable runtime introspection.
\subsection{Memory Analysis of Python}
\label{s2_3}
Python’s runtime architecture consists of interdependent memory management subsystems that control object allocation and persistence in memory. The global \texttt{\_PyRuntime} structure, of type \texttt{\_PyRuntimeState}, represents the entry point to the runtime, maintaining interpreter registries, thread states, and the configuration of the garbage collector. This structure is located in the dynamic symbol table of \textit{ELF} binaries on Linux or the export table of \textit{PE} binaries on Windows \citep{ali2025leveraging}. Each interpreter instance (\texttt{PyInterpreterState}) maintains its own registry, import machinery, and a set of thread contexts (\texttt{PyThreadState}), where each thread holds a stack of execution frames containing references to active function objects \citep{ali2025memory}. Memory allocation follows a hybrid scheme.  Small objects ($\leq$512 bytes) are managed by the \texttt{pymalloc} allocator within 256 KB arenas subdivided into 4 KB pools \citep{arena}. Larger objects are handled by the system allocator via \texttt{brk()} for the process heap and \texttt{mmap()} for separate anonymous memory mappings, depending on allocation size and system configuration. The garbage collector tracks objects across three generations using doubly-linked lists via \texttt{gc\_prev} and \texttt{gc\_next} pointers. These mechanisms define the spatial organization of Python objects in memory, enabling the reconstruction of complete runtime SBOMs from memory artifacts.

\section{Related Work}
\label{s3}
\textbf{Evaluating Existing SBOM Tools.}
Prior studies primarily evaluated SBOM tools, standardization challenges, and adoption barriers.
Dalia \textit{et al.} \citep{dalia2024sbom}  evaluated \textit{Syft}, \textit{Microsoft SBOM Tool}, \textit{Tern}, and \textit{CycloneDX} tools in terms of multi-platform support, format compatibility, and usability, demonstrating that these tools remain static, metadata-driven, and lacking operational readiness.
Mirakhorli \textit{et al.} \citep{mirakhorli2024landscape} analyzed 84 tools and observed heavy reliance on package managers, inconsistent component identification, and limited cross-language coverage, urging the need for standardized dependency semantics and validation frameworks.
Halbritter and Merli \citep{halbritter2024accuracy} further showed that even the most accurate tools (\textit{CycloneDX-Npm} and \textit{CycloneDX-Conan}) fail to meet the \textit{NTIA}'s minimum-element requirements, frequently omitting dependencies or misreporting versions. From an adoption perspective, Xia \textit{et al.} \citep{xia2023empirical}  found that standardization issues, inadequate validation, and information-sensitivity concerns hinder adoption. Stalnaker \textit{et al.} \citep{stalnaker2024boms} identified specification complexity, tooling insufficiency, and verification issues as major barriers, and highlighted the need for language-specific and verifiable SBOM frameworks, particularly for emerging AI/ML domains lacking formal standards.  Bi \textit{et al.} \citep{bi2024way} analyzed 4,786 SBOM-related GitHub discussions, linking poor SBOM quality to technical debt and governance deficiencies, highlighting the need for continuous assurance integration.  Complementary studies by Cofano \textit{et al.} \citep{cofano2024sbom} and Yu \textit{et al.} \citep{yu2024correctness} highlighted consistent reliability issues in \textit{Trivy}, \textit{Syft}, \textit{CDXGen}, \textit{ORT}, \textit{Microsoft SBOM Tool}, and \textit{GitHub Dependency Graph}, attributing inaccuracies to flawed metadata parsing, ambiguous naming and version-resolution, and incomplete dependency resolution, reflecting the inherent limitations of static, metadata-driven analysis.

\textbf{Proposing New SBOM Techniques.}
Recent efforts have extended SBOM generation to the deployment phase of software.
Sharma \textit{et al.} \citep{sharma2024sbom} introduce \textit{SBOM.EXE}, which prevents dynamic code injection in Java by maintaining a Bill of Materials Index (BOMI) of legitimate class checksums. This approach is inherently limited since it requires complete test coverage and cannot handle non-deterministic code generation or hidden classes.
Benedetti \textit{et al.} \citep{benedetti2024impact}  propose \textit{PIP-SBOM}, which integrates SBOM creation into the Python package manager \textit{pip} using its native resolver.  However, this approach reports versions that would be freshly installed rather than actual deployed versions, and inherits pip's limitations in failing to capture runtime-only dependencies and dynamic imports. Kawaguchi and Hart \citep{kawaguchi2024deployment} developed \textit{C3DRS}, a consumer-side vulnerability management system that generates SBOMs through dynamic container inspection. Although it improves detection accuracy, its analysis remains limited to filesystem artifacts.
\section{Challenges}
\label{s4}
This section identifies six challenges that \textsc{MEM-SBOM} is designed to overcome.

\noindent \textbf{C1: Evasive manipulation of the Python import system.}
The dynamic nature of Python allows introspection and modification of its import system during execution. The same flexibility that enables extensibility, via monkey patching, custom import hooks, or dynamic loaders, can be exploited to conceal or alter modules in memory. Attackers may delete or overwrite entries in the interpreter registry (\texttt{sys.modules}), replace legitimate modules with malicious substitutes, create modules that never appear in \texttt{sys.modules} using the \texttt{importlib} and  \texttt{runpy} APIs, inject hidden modules into sub-interpreters, or unlink module objects from the garbage collector by manipulating their pointers directly through \texttt{ctypes}.  Such behaviors complicate the recovery of a complete and trustworthy module set from memory.
\noindent \textbf{C2: Fragmented runtime state in multi-process applications.}
Complex Python applications (e.g., \textit{Airflow}, \textit{FastAPI}, \textit{MLflow}) comprise multiple cooperating processes, each maintaining its own interpreter state and dependency set. These processes may load different or conflicting versions of shared packages depending on the deployment environment. To generate an accurate, application-level SBOM, these fragmented module views must be merged while resolving version conflicts and cross-process duplications.
\noindent \textbf{C3: Distinguishing third-party packages from internal and standard libraries.}
When a Python application runs, the interpreter instantiates not only the application’s modules but also a wide range of built-in and standard-library modules required for its own operations. In memory, all of these appear as \texttt{PyModuleObject} instances within the same runtime, making it difficult to isolate third-party components of the application. Differentiating these modules is essential for constructing an accurate SBOM that reflects only third-party dependencies.
\noindent \textbf{C4: Incomplete version information in memory.}
Accurately resolving package versions from memory poses a major challenge for memory-based SBOM generation. A \texttt{PyModuleObject} may expose version-related attributes, such as \texttt{\_\_version\_\_}, \texttt{VERSION}, \texttt{\_\_VERSION\_\_}, \texttt{version\_info},  \texttt{\_\_version\_info\_\_}. However, these attributes lack standardization, exhibit inconsistent naming, and are not guaranteed to be present in memory. This challenge complicates the generation of accurate memory-based SBOMs, for which package name and version are essential elements.
\noindent \textbf{C5: Structural over-approximation of dependency relationships.}
Python’s eager import mechanism loads entire namespaces into memory, making it difficult to distinguish between modules that are merely loaded and those actually used. When a module imports another, the callee’s modules, classes, and functions are instantiated in memory, even if most remain unreferenced. Recursive imports further amplify this effect by pulling additional transitive dependencies into memory. Traversing these imports results in deep transitive chains that populate memory with modules the application never invokes, leading to inflated SBOMs with high false-positive rates. This, in turn, causes the dependency graph to over-approximate the true runtime state, reporting relationships that are syntactically derived but semantically invalid. Therefore, with such structural over-approximation, constructing precise runtime SBOMs poses another challenge.
\noindent \textbf{C6: Identifying exploitable code paths.}
Beyond C5’s module-level over-approximation, this challenge extends to the internal objects of each module. Once imported, all of a module’s objects, such as functions, classes, and attributes, reside in memory, even though many of them may never be used by the application. Thus, the presence of a vulnerable function in memory does not imply that it is reachable from the application. Determining which portions of a module are affected by a vulnerability requires inspecting all of its objects and enumerating the possible call chains across all application modules, which becomes a core technical challenge.

\section{Methodology}
\label{s5}
\begin{figure*}[!htb]
\centering
 \includegraphics[width=\textwidth]{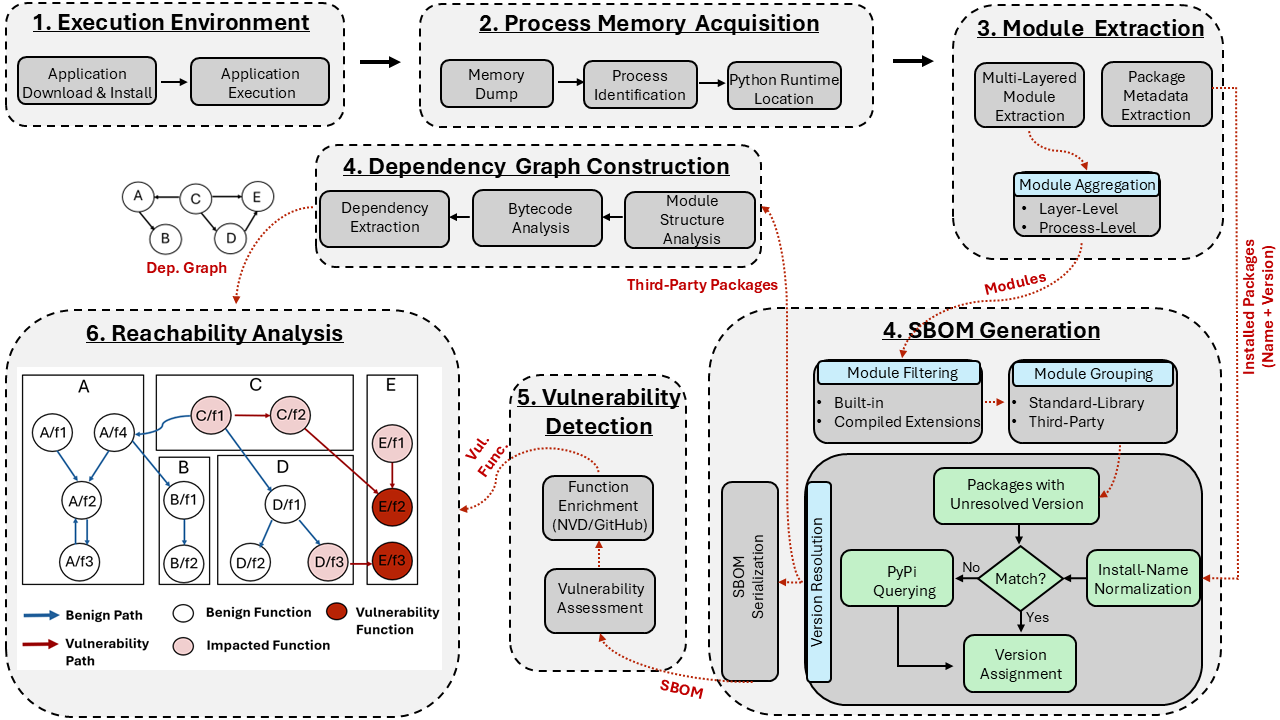}
\caption{Overview of the \textsc{MEM-SBOM} Framework}
 \label{meth_1}
\vspace{-0.1in}
\end{figure*}

Figure \ref{meth_1} illustrates the \textsc{MEM-SBOM} framework. After identifying the application processes and locating the interpreter's internal structures within the memory dump \citep{ali2025memory}, the framework extracts the runtime modules to produce a CycloneDX-compliant SBOM, a dependency graph, and function-level exploitation paths.
\subsection{Module Extraction}
\label{s5_1}
\subsubsection{Package Metadata Extraction}
\label{s5_1_1}
To identify installed packages without filesystem access, \textsc{MEM-SBOM} inspects the Python import system cache (i.e., \texttt{sys.path\_importer\_cache}), which maps filesystem paths to finder objects responsible for module discovery (see Section~\ref{s2_2}). Each \texttt{FileFinder} instance maintains a persistent \texttt{\_path\_cache} containing all recognized directory entries, including source files (\texttt{.py}), bytecode files (\texttt{.pyc}), extension modules (\texttt{.so}, \texttt{.pyd}), and package-metadata directories (\texttt{.dist-info}, \texttt{.egg-info}). This cache is constructed once at import time, providing a stable record of the interpreter’s package environment. \textsc{MEM-SBOM} extracts the package metadata (i.e., name and version) from cached directory entries that match the \textit{PEP 427} wheel format (\texttt{package\_name-version.dist-info}) or legacy \texttt{egg-info} naming scheme. It then normalizes the recovered names, associates version numbers, and identifies their installation scope based on the directory paths. 

\subsubsection{Multi-Layered Module Extraction}
\label{s5_1_2}
Python’s runtime structures can be exploited to conceal malicious modules through various evasion techniques, as discussed next.

\textbf{A1. Module Deletion.} Deletes entries from \texttt{sys.modules}, hiding modules that remain live through residual references elsewhere in memory.  
\textbf{A2. Module Overwrite.} Overwrites or aliases legitimate entries in \texttt{sys.modules} (e.g., mapping a benign name to a malicious module object) to hide the module behind a trusted identifier.
\textbf{A3. Import-Hook Interception.} Inserts a malicious finder into \texttt{sys.meta\_path} that redirects a targeted module import to a custom loader, which executes a malicious module during initialization.
\textbf{A4. Import-Bypassing Module Creation.} Creates modules via \texttt{importlib} and \texttt{runpy} APIs without registering them in \texttt{sys.modules}.
\textbf{A5. Sub-Interpreter Injection.} Injects malicious modules into secondary interpreters (sub-interpreters) within the same process, making them invisible to scans of the main interpreter's registry.
\textbf{A6. Garbage-Collector (GC) Manipulation.} Manipulates the pointers of GC-tracked objects (e.g., via \texttt{ctypes}) to unlink malicious modules from its doubly-linked lists, allowing them to evade detection.  

To address these evasion techniques, \textsc{MEM-SBOM} employs a robust multi-layered extraction approach that overcomes Challenge \textbf{C1} by incrementally expanding coverage from interpreter registries to raw heap regions.  Layers 1–3 provide authoritative and context-rich evidence with minimal false positives,  ideal for benign module recovery. Layers 4–5 extend extraction to raw memory arenas and heap regions. Although these layers require object-type validation, they are essential for achieving completeness, as all modules with active references remain allocated in memory.
\textbf{L1. Interpreter Registry.} Traverses \texttt{sys.modules} of each interpreter, providing the fastest extraction mechanism. 
\textbf{L2. Thread Execution States.} Inspects the global namespace and local variables of the stack frames to identify their modules that might be hidden from the interpreter registry.
\textbf{L3. Garbage Collector Traversal.} Traverses the doubly-linked lists of objects managed by the process-level garbage collector, which is shared across all interpreters.
\textbf{L4. Arena-Aware Analysis.} Scans arena pools to identify valid module objects within allocated blocks by verifying their reference count (\texttt{ob\_refcnt}) and type (\texttt{ob\_type}). 
\textbf{L5. Heap Areas Scanning.} Recovers large module objects in 8-byte aligned increments, interpreting 16-byte sequences as \texttt{ob\_refcnt} and \texttt{ob\_type}, validating the \texttt{PyObject} header, dereferencing the type pointer if it corresponds to a module type, and confirming module identity. 

\noindent\textbf{Module Aggregation.}
To address Challenge \textbf{C2}, \textsc{MEM-SBOM} implements two levels of aggregation.  First, within each application process, it deduplicates modules discovered across all five extraction layers by name. Second, for multi-process applications, per-process module sets are merged to generate the application's complete module set. 

\subsection{SBOM Generation}
\label{s5_2}
After module extraction, \textsc{MEM-SBOM} retains only the application packages for version resolution, excluding built-in and standard-library components from the SBOM. These packages are then serialized into a CycloneDX-compliant SBOM for further analysis.

\subsubsection{Module Filtering}
\label{s5_2_1}
A Python process includes the application’s own modules, standard-library modules, and a set of built-in modules that are compiled directly into the interpreter. The application itself is structured as a main package with its dependent packages, each composed of one or more \texttt{.py} files represented in memory as \texttt{PyModuleObject} instances \citep{pep451, pep420}.  
The main package corresponds to the distribution’s top-level package (e.g., \textit{streamlit}, \textit{Django}, \textit{Pelican}) while all other imported packages are treated as dependencies. Addressing Challenge \textbf{C3}, \textsc{MEM-SBOM} first filters out the built-in modules (e.g., \texttt{\_signal}, \texttt{marshal}, and \texttt{\_imp}) that are identified through the interpreter’s \texttt{sys.builtin\_module\_names} registry. It further excludes the compiled extension modules (e.g., \texttt{.so}, \texttt{.pyd}) that represent native C extensions. The remaining modules (i.e., standard-library and third-party) are retained for grouping and top-level package mapping in the next stage.

\subsubsection{Module Grouping}
\label{s5_2_2}
Each module is identified by its fully qualified name (FQN), which specifies its position within the package hierarchy. For example, the package \texttt{requests} with its modules (\texttt{adapters}, \texttt{models}, and \texttt{sessions}) are represented in memory as  \texttt{requests}, \texttt{requests.adapters}, \texttt{requests.models}, and \texttt{requests.sessions}. \textsc{MEM-SBOM} groups modules by their top-level package, defined as the substring preceding the first dot (\texttt{.}) in the module name. This grouping recovers a one-to-many relationship between a package and the modules that compose it, enabling accurate package-level representation in the resulting SBOM. Among these top-level packages, the application's main package appears as one of them and is grouped in the same way as all other packages.
After grouping, the framework classifies each package as standard-library or third-party based on its installation path and excludes the standard-library packages. Third-party packages are identified by paths such as \texttt{site-packages}, \texttt{dist-packages}, or environment-specific vendor directories (e.g., \texttt{vendor-packages}, \texttt{local-packages}). Standard-library packages are recognized through name-based heuristics and path patterns corresponding to the interpreter’s installation directories (e.g., \texttt{/usr/lib/python3.x/}, \texttt{/lib/python3.x/}).

Some packaging tools bundle their own dependencies to ensure version consistency, creating namespace-isolated copies of third-party packages. For instance, \textit{pip} embeds its dependencies under \texttt{pip.\_vendor} (e.g., \texttt{pip.\_vendor.requests}), while \textit{setuptools} uses \texttt{setuptools.\_vendor}.  \textsc{MEM-SBOM} identifies and excludes these bundled copies by matching common namespace conventions (e.g., \texttt{.\_vendor.}, \texttt{.extern.}) and directory patterns (e.g., \texttt{/\_vendor/}, \texttt{/extern/}).  Internal utility packages, such as \texttt{\_distutils\_hack}, are similarly omitted to ensure the SBOM reflects only the application packages.

\subsubsection{Version Resolution}
\label{s5_2_3}
To identify the version of the resulting packages from the previous stage, \textsc{MEM-SBOM} analyzes their \texttt{PyModuleObject} structures to inspect their version attributes (e.g., \texttt{\_\_version\_\_}, \texttt{VERSION}, \texttt{version\_info}). As noted in Challenge \textbf{C4}, these attributes are not mandatory and therefore might not appear in memory. When version fields cannot be located, the framework attempts to map each import name to its installation record in \texttt{sys.path\_importer\_cache} to retrieve the associated version (see Section \ref{s5_1_1}).  However, distribution names in installation records differ from their import counterparts due to the normalization rules defined in \textit{PEP 503} \citep{pep503}. For example, the distribution \texttt{email-validator} corresponds to the import name \texttt{email\_validator}.  Such discrepancies are resolved by applying \textit{PEP 503} normalization (i.e., case folding and hyphen-to-underscore transformation) before matching import and installation names.

A more complex case arises when normalization fails to establish a correspondence between the import names observed in memory and the installation names in the local cache (\texttt{sys.path\_importer\_cache}). For instance, \texttt{sys.modules} may contain \texttt{cv2}, while the cache lists \texttt{opencv-contrib-python}, and \texttt{opencv-python-headless}. Although these distributions originate from the same upstream \textit{OpenCV} project, they install distinct packages that all expose the same import name (\texttt{cv2}), making it impossible to infer from the local cache alone which distribution provided the loaded module. 

To resolve such ambiguity, and given that the PyPI API operates on distribution identifiers rather than import names, \textsc{MEM-SBOM} queries PyPI using the previously unmatched installation names discovered in the local cache. The framework retrieves package metadata through the public JSON API, extracting import names from the \texttt{.dist-info/top\_level.txt} file for wheel distributions and from the \texttt{.egg-info/top\_level.txt} file for source distributions. This file enumerates the top-level importable names defined by each distribution. 
This process allows the recovery ofing the complete set of import names corresponding to the locally installed distributions. \textsc{MEM-SBOM} then matches these recovered import names against the modules observed in memory to determine their originating distributions. Once a match is found, the version information is obtained from the local cache, ensuring that the generated SBOM reflects the actual environment state. Ultimately, when the version attribute remains indeterminate, \textsc{MEM-SBOM} records its value as \texttt{"unknown"}, consistent with CISA’s SBOM minimum-element specification~\citep{cisa2025sbom_1}.

\subsubsection{SBOM Serialization}
\label{s5_2_4}
After resolving package versions, \textsc{MEM-SBOM} serializes the package information into a CycloneDX-compliant SBOM in JSON format. We selected this standard due to its security-oriented design and support for vulnerability correlation, which aligns with the cybersecurity and digital forensics support objectives of our work. Following CISA’s guidance \citep{cisa2024sbom_2}, we focus on attributes essential to incident response (e.g.,  component identity, version, and dependency relationships)  rather than licensing or copyright metadata.
The generated SBOM consists of four primary sections.
\textbf{Metadata.} Defines SBOM provenance using a UUID-based serial number, extraction timestamp, and tool identifier to ensure extraction traceability.
\textbf{Components.} Lists each package with fields \texttt{name}, \texttt{version}, and package URL (PURL) following (\texttt{pkg:pypi/package@version}) format, enabling direct mapping to vulnerability databases such as National Vulnerability Database (NVD), Open Source Vulnerabilities (OSV), and GitHub Security Advisories (GHSA). Component hashes are omitted because memory-resident modules lack stable file artifacts, consistent with CISA’s allowance for inaccessible sources. Additional forensic metadata, such as \texttt{extracted\_from: memory}  to distinguish runtime from manifest-based SBOMs, and \texttt{file\_path} for filesystem correlation, are preserved in the \texttt{properties} of the component.
\textbf{Dependencies.}  Lists the direct dependencies of the application packages.
\textbf{Annotations.} Includes the annotator identity, creation timestamp, extraction source, and package count, allowing analysts to assess completeness and maintain provenance across multiple SBOMs.

\subsection{Dependency Graph Construction}
\label{s5_3}
This section addresses Challenge \textbf{C5} by describing how \textsc{MEM-SBOM} constructs the dependency graph that captures the relationships among the recovered application packages.

\subsubsection{Module Structure Analysis}
\label{s5_3_1}
\begin{sloppypar}
The namespace dictionary (\texttt{\_\_dict\_\_}) of each \texttt{PyModuleObject} contains references to the module’s own objects as well as those imported from other modules. For module objects, the defining module is identified through the \texttt{md\_name} attribute; for class objects, through the \texttt{\_\_module\_\_} attribute; and for callable objects (e.g., functions, generators, coroutines, asynchronous generators, instance methods, class methods, and static methods), through the \texttt{func\_module} attribute. The analysis is applied recursively to capture dependencies within nested classes and inner functions. 
\end{sloppypar}
\subsubsection{Bytecode Analysis}
\label{s5_3_2}
To extract import statements embedded within the function body, we disassemble its compiled bytecode and apply a stride-1 sliding window of width four over the opcode sequence to capture the complete instruction patterns associated with import variants. During traversal, non-semantic scaffolding instructions (e.g., \texttt{RESUME}, prologue assignments, and constant initializations) are excluded to reduce noise. Each window is then compared against the opcode patterns summarized in Table \ref{bytecode_analysis_1}.

\textbf{Standard Import Patterns (P1-P3).}  Direct import statements (\texttt{import module})  generate an \texttt{IMPORT\_NAME} instruction, followed by a scope-dependent storage instruction (\texttt{STORE\_FAST} for function scope or \texttt{STORE\_NAME} for module/class scope). Aliased imports yield equivalent opcode sequences repeated per module, with the module name extracted from \texttt{IMPORT\_NAME.argval} field.

\textbf{From-Import Patterns (P4-P8).} The \texttt{from module import item} statement generates an \texttt{IMPORT\_NAME} followed by one or more \texttt{IMPORT\_FROM} instructions and concluding with \texttt{POP\_TOP}. The analysis extracts only the base module name, as imported symbols do not influence dependency relationships. The star-import variant (\texttt{from module import *}) replaces \texttt{IMPORT\_FROM} with \texttt{IMPORT\_STAR} and omits \texttt{POP\_TOP}.

\textbf{Dynamic Import Patterns (P9-P11).} Function-based imports, called via \texttt{exec()}, \texttt{importlib.import\_module()}, or \texttt{\_\_import\_\_()}, are detected by recognizing their characteristic opcode call sequences rather than dedicated import opcodes. For \texttt{exec()}, the analysis inspects the \texttt{LOAD\_CONST} string operand and applies regular expressions to extract embedded import statements. In contrast, \texttt{importlib} and \texttt{\_\_import\_\_()} embed the target module name as a string constant within the \texttt{LOAD\_CONST} instruction.

\begin{sloppypar}
\textbf{Note.} The bytecode analysis maintains cross-version compatibility by handling instruction-set variations across Python releases. Bound-method calls compiled as \texttt{LOAD\_METHOD} $\rightarrow$  \texttt{CALL\_METHOD} in Python 3.6–3.10 are replaced in Python 3.11 by \texttt{LOAD\_ATTR}$\rightarrow$ \texttt{CALL}, optionally followed by \texttt{KW\_NAMES} to handle keyword arguments. For name resolution, module-level code uses \texttt{LOAD\_NAME}$\rightarrow$  \texttt{STORE\_NAME}, while function scopes use \texttt{LOAD\_GLOBAL}$\rightarrow$ \texttt{STORE\_GLOBAL} for global variables, \texttt{LOAD\_FAST} $\rightarrow$  \texttt{STORE\_FAST} for local variables, and \texttt{LOAD\_DEREF} $\rightarrow$  \texttt{STORE\_DEREF} for closure variables. When \texttt{LOAD\_CONST} instructions have a code object as their argument value, the analysis recursively inspects them to recover imports from all inner scopes.
\end{sloppypar}
\begin{table*}[!htbb]
\small
\centering
\caption{Bytecode patterns corresponding to Python import constructs.}
\label{bytecode_analysis_1}
\begin{tabular}{p{3cm}|p{4.9cm}|p{8cm}}
\toprule
\textbf{Pattern} & \textbf{Python Code Example} & \textbf{Bytecode Sequence Pattern}\\
\midrule
P1: Direct Import 
& \texttt{import math} 
& \texttt{IMPORT\_NAME(math)} $\rightarrow$ \texttt{STORE\_FAST(math)} \texttt{/} \texttt{STORE\_NAME(math)} \\
\midrule

P2: Multiple Direct Imports 
& \texttt{import os, sys} 
& Repeat: (\texttt{IMPORT\_NAME} $\rightarrow$ \texttt{STORE\_FAST} \texttt{/} \texttt{STORE\_NAME}) for each module\\
\midrule

P3: Direct Import with Alias 
& \texttt{import math as m} 
& \texttt{IMPORT\_NAME(math)} $\rightarrow$ \texttt{STORE\_FAST(m)}\texttt{/} \texttt{STORE\_NAME(m)}\\
\midrule

P4: From Import  \newline (Single Module) 
& \texttt{from math import sqrt} 
& \texttt{IMPORT\_NAME(math)} $\rightarrow$ \texttt{IMPORT\_FROM(sqrt)} $\rightarrow$ \texttt{STORE\_FAST(sqrt)}\texttt{/} \texttt{STORE\_NAME(sqrt)} $\rightarrow$ \texttt{POP\_TOP}\\
\midrule

P5: From Import (Multiple Modules) 
& \texttt{from math import sqrt, pi, cos}
& \texttt{IMPORT\_NAME(math)} $\rightarrow$ Repeat: (\texttt{IMPORT\_FROM} $\rightarrow$ \texttt{STORE\_FAST} \texttt{/} \texttt{STORE\_NAME})  for each item $\rightarrow$ \texttt{POP\_TOP}\\
\midrule

P6: From Import with Alias 
& \texttt{from math import sqrt as sq}
& \texttt{IMPORT\_NAME(math)} $\rightarrow$ \texttt{IMPORT\_FROM(sqrt)} $\rightarrow$ \texttt{STORE\_FAST(sq)} \texttt{/} \texttt{STORE\_NAME(sq)} $\rightarrow$ \texttt{POP\_TOP}\\
\midrule

P7: From Import Star 
& \texttt{from math import *} 
& \texttt{IMPORT\_NAME(math)} $\rightarrow$ \texttt{IMPORT\_STAR}\\
\midrule

P8: From Import (Submodules) 
& \texttt{from os.path import join} 
& \texttt{IMPORT\_NAME(os.path)} $\rightarrow$ \texttt{IMPORT\_FROM(join)} $\rightarrow$ \texttt{STORE\_FAST(join)} \texttt{/} \texttt{STORE\_NAME(join)} $\rightarrow$ \texttt{POP\_TOP}\\
\midrule

P9: Dynamic Import via \texttt{exec()}
& \texttt{exec('import math')} 
& \texttt{LOAD\_GLOBAL/LOAD\_NAME}(exec) $\rightarrow$ \texttt{LOAD\_CONST(string)} $\rightarrow$ \texttt{CALL\_FUNCTION/CALL}\\
\midrule

P10: Dynamic Import via \texttt{importlib} 
& \texttt{import importlib} 
  \texttt{m = importlib.} \newline
  \texttt{~~import\_module('math')}
& \texttt{IMPORT\_NAME(importlib)} $\rightarrow$ ... $\rightarrow$ \texttt{STORE\_FAST(importlib)} \texttt{/} \texttt{STORE\_NAME(importlib)} $\rightarrow$ \texttt{LOAD\_METHOD/LOAD\_ATTR(import\_module)} $\rightarrow$ \texttt{LOAD\_CONST('math')} $\rightarrow$ \texttt{CALL\_METHOD/CALL}\\
\midrule

P11: Dynamic Import via \texttt{\_\_import\_\_()} 
& \texttt{m = \_\_import\_\_('math')}
& \texttt{LOAD\_GLOBAL/LOAD\_NAME(\_\_import\_\_)} $\rightarrow$ \texttt{LOAD\_CONST('math')} $\rightarrow$ \texttt{CALL\_FUNCTION/CALL}\\
\bottomrule
\end{tabular}
\end{table*}

\subsubsection{Dependency Extraction}
\label{s5_3_3}
Algorithm \ref{alg1} constructs the dependency graph by integrating namespace analysis and bytecode inspection. For each application package $p \in \mathcal{P}$, let $\mathcal{M}_p$  denote the set of modules belonging to $p$. For each module $m \in \mathcal{M}_p$, the algorithm initializes an empty import set $\mathcal{I}_{m}$ and traverses the module's dictionary. For each object $o$ in the dictionary, it determines the object type $\tau$ through the \texttt{ob\_type} field. When $\tau$ is \texttt{module} and the referenced module name belongs to a different package than $p$, it is identified as an external dependency and appended to $\mathcal{I}_{m}$. For class objects, it extracts the \texttt{\_\_module\_\_} attribute and recursively processes the class dictionary to capture nested dependencies. For callable objects, it extracts the \texttt{func\_module} reference,  validates it as an external dependency, analyzes its bytecode, and merges the results into $\mathcal{I}_{m}$.
After computing $\mathcal{I}_{m}$ for all modules of package $p$, the algorithm aggregates and deduplicates imports across $\mathcal{M}_p$, and creates a directed edge $E(p, q)$ for each distinct imported package $q$, yielding a package-level dependency graph.

\begin{algorithm}[!htb]
\caption{\textit{\textbf{Dependency Graph Construction}}}
\begin{algorithmic}[1]
\STATE \textbf{Input:} $\mathcal{P}$ (application packages)
\STATE \textbf{Output:} $\mathcal{D}_{graph}$ (dependency graph)
\STATE Initialize $\mathcal{D}_{graph} \gets \emptyset$
\FOR{each package $p \in \mathcal{P}$}
    \STATE Initialize $\mathcal{I}_{p} \gets \emptyset$ \COMMENT{Imports observed for package $p$}
    \FOR{each module $m \in \mathcal{M}_p$}
        \STATE Initialize $\mathcal{I}_{m} \gets \emptyset$
        \STATE \textsc{ProcessDict}$(p, m.\_\_dict\_\_, \mathcal{I}_{m})$
        \STATE $\mathcal{I}_{p} \gets \mathcal{I}_{p} \cup \mathcal{I}_{m}$
    \ENDFOR
    \FOR{each imported package $q \in \mathcal{I}_{p}$}
        \STATE $\mathcal{D}_{graph} \gets \mathcal{D}_{graph} \cup \{E(p, q)\}$
    \ENDFOR
\ENDFOR
\STATE
\STATE \textbf{Procedure} \textsc{ProcessDict}$(p, dict, \mathcal{I}_{m})$
\FOR{each object $o \in dict$}
    \STATE $\tau \gets o.\text{ob\_type}.\text{tp\_name}$
    \IF{$\tau = \text{module}$}
        \STATE $mod\_name \gets o.\_\_name\_\_$
        \IF{$pkg(mod\_name) \neq p$}
            \STATE $\mathcal{I}_{m} \gets \mathcal{I}_{m} \cup \{mod\_name\}$
        \ENDIF
    \ELSIF{$\tau = \text{class}$}
        \STATE $class\_mod \gets o.\_\_module\_\_$
        \IF{$pkg(class\_mod) \neq p$}
            \STATE $\mathcal{I}_{m} \gets \mathcal{I}_{m} \cup \{class\_mod\}$
        \ENDIF
        \STATE \textsc{ProcessDict}$(p, o.\_\_dict\_\_, \mathcal{I}_{m})$ \COMMENT{Recurse into class}
    \ELSIF{$\tau \in Callable\_Types$}
        \STATE $func\_mod \gets o.\text{func\_module}$
        \IF{ $pkg(func\_mod) \neq p$}
            \STATE $\mathcal{I}_{m} \gets \mathcal{I}_{m} \cup \{func\_mod\}$
        \ENDIF
        \STATE $\mathcal{I}_{bytecode} \gets$ \textsc{BytecodeAnalysis}$(o.\text{func\_code})$
        \STATE $\mathcal{I}_{m} \gets \mathcal{I}_{m} \cup \mathcal{I}_{bytecode}$
    \ENDIF
\ENDFOR
\end{algorithmic}
\label{alg1}
\end{algorithm}


\subsection{Vulnerability Detection}
\label{s5_4}
To address Challenge \textbf{C6}, \textsc{MEM-SBOM} integrates the SBOM, dependency graph, and public vulnerability data to identify vulnerable packages within the application, determine their impact scope, and trace all reachable paths to them. To identify vulnerabilities, we use \textit{Grype} (v0.100.0) as the vulnerability scanner due to its accuracy, ecosystem coverage, and competitive performance compared to alternative scanners \citep{grype, benedetti2024impact, bufalino2025sbomproof}. This tool maps SBOM components (via their PURLs) to known CVEs using aggregated advisory feeds (e.g., NVD, OSV, and GitHub Security Advisories). For each SBOM component, it produces a structured set of matching vulnerabilities, including identifiers, affected version ranges, and severities, which \textsc{MEM-SBOM} attaches to the corresponding packages as input to the function-level analysis.

\subsubsection{Function-Level Enrichment}
\label{s5_4_1}
Given the over-approximation inherent in package-level vulnerability detection, as discussed in Challenges \textbf{C5} and \textbf{C6}, \textsc{MEM-SBOM} augments CVE matches with function-level metadata extracted from public vulnerability advisories. This enrichment enables us to distinguish vulnerable code paths within the application. For each CVE reported by \textit{Grype}, we query NVD and GitHub Security Advisories to obtain additional context. From NVD, we retain entries whose CPE configurations match the affected package and apply heuristics over the free-text CVE description to extract function identifiers. From GitHub advisories, when remediation commits or pull requests are available, we analyze the corresponding diffs to identify updated function definitions. When function-level attribution cannot be reliably determined, either due to incomplete advisory descriptions or genuinely package-wide impact, we label the CVE with package-wide scope and leave finer-grained localization to manual analysis.

\subsection{Reachability Analysis}
\label{s5_5}
\textsc{MEM-SBOM} determines which parts of the application are exposed to identified CVEs by tracing exploitation paths through backward taint propagation. Given a vulnerability sink $\mathcal{S} = (p, m, f)$, where $p$ denotes the vulnerable package, $m$ the module, and $f$ the function, it traces backward from $f$ to identify all callers that eventually reach the application's main package. This produces taint propagation paths $\langle f_1, f_2, \ldots, f_n \rangle$, where $f_n = f$ is the vulnerable function and $f_1$ belongs to the application's main package. The sink is considered reachable if at least one such path exists. 

The analysis proceeds in two stages. First, within the vulnerable package, we inspect all modules, classes, and functions to identify local callers of $f$. Any function that invokes $f$, either directly or through an internal call chain within the same package, is added to the sink set. Second, for each package that imports the vulnerable package, we detect functions that directly call any element of the current sink set and mark them as vulnerable. We then apply the same intra-package analysis to these caller packages to identify internal call chains that reach those functions. The resulting set of direct and internal callers forms the sink set for the next iteration. This recursive expansion continues until the process reaches the application's main package. At that point, we perform the same intra-package analysis on the main package to enumerate all application functions involved in the call chain. A vulnerability is considered reachable if any backward traversal originating in the main package reaches the vulnerable function.

This process requires analyzing function bytecode to extract call relationships. Therefore, we extend the bytecode analysis introduced in Section \ref{s5_3_2} to detect the call patterns shown in Table \ref{bytecode_analysis_2}, including inner-function calls, closure and nonlocal calls, decorator-wrapper patterns, and instance, class, and module-level function calls.

\begin{table*}[!htb]
\small
\centering
\caption{ Bytecode patterns corresponding to Python function call constructs.}
\label{bytecode_analysis_2}
\begin{tabular}{p{3cm}|p{4.9cm}|p{8cm}}
\toprule
\textbf{Pattern} & \textbf{Python Code Example} & \textbf{Bytecode Sequence Pattern}\\
\midrule

F1: Inner Function Call
& \texttt{def outer():} \newline \texttt{~~def inner(): ...} \newline \texttt{~~inner()}
& \textbf{Definition:} \texttt{LOAD\_CONST(inner code object)} $\rightarrow$ \texttt{MAKE\_FUNCTION} $\rightarrow$ \texttt{STORE\_FAST(inner)} \newline \textbf{Call:} \texttt{LOAD\_FAST(inner)} $\rightarrow$ \texttt{CALL\_FUNCTION} \texttt{/} \texttt{CALL} \texttt{/} \texttt{CALL\_FUNCTION\_KW} \texttt{/} \texttt{CALL\_FUNCTION\_EX}  \\
\midrule

F2: Closure Function Call
& \texttt{def outer():} \newline \texttt{~~x = 10} \newline \texttt{~~def inner():} \newline \texttt{~~~~return x} \newline \texttt{~~inner()}
& \textbf{Definition:} \texttt{LOAD\_CLOSURE(x)}  $\rightarrow$ \texttt{LOAD\_CONST(inner code object)} $\rightarrow$ \texttt{MAKE\_FUNCTION(closure)} $\rightarrow$ \texttt{STORE\_FAST(inner)} \newline \textbf{Call:} \texttt{LOAD\_FAST(inner)} $\rightarrow$ \texttt{CALL\_FUNCTION} \texttt{/} \texttt{CALL} \texttt{/} \texttt{CALL\_FUNCTION\_KW} \texttt{/} \texttt{CALL\_FUNCTION\_EX}  \\
\midrule

F3: Closure/Nonlocal Call
& \texttt{def outer():} \newline \texttt{~~def helper(): ...} \newline \texttt{~~def inner():} \newline \texttt{~~~~helper()}
& \texttt{LOAD\_DEREF(helper)} $\rightarrow$ \texttt{CALL\_FUNCTION} \texttt{/} \texttt{CALL} \texttt{/} \texttt{CALL\_FUNCTION\_KW} \texttt{/} \texttt{CALL\_FUNCTION\_EX} \\
\midrule

F4: Decorator Wrapper Pattern Call
& \texttt{def wrapper(*args, **kwargs):} \newline  \texttt{~~return func(*args, **kwargs)}
& \textbf{Definition:} \texttt{LOAD\_CONST(inner code object)} $\rightarrow$ \texttt{MAKE\_FUNCTION}(closure) $\rightarrow$ \texttt{STORE\_FAST(wrapper)} \newline
\textbf{Inside wrapper - calling original:} \texttt{LOAD\_DEREF(func)} $\rightarrow$  \texttt{LOAD\_FAST(args)} $\rightarrow$ \texttt{LOAD\_FAST(kwargs)} $\rightarrow$ \texttt{CALL\_FUNCTION\_EX}\\
\midrule

F5: Function Call (instance/class/module) & \texttt{obj = MyClass()},  \texttt{obj.method()}  \newline \texttt{module.func()}  & \texttt{LOAD\_METHOD()} \texttt{/} \texttt{LOAD\_ATTR()} $\rightarrow$  ...  $\rightarrow$ \texttt{CALL\_METHOD}\texttt{/} \texttt{CALL}  \texttt{/} \texttt{CALL\_FUNCTION\_KW} \texttt{/} \texttt{CALL\_FUNCTION\_EX} \\
\midrule
 \end{tabular}
\end{table*}
\section{Evaluation}
\label{s6}
This section first validates the capabilities of \textsc{MEM-SBOM} in generating accurate runtime SBOMs across heterogeneous installation environments, constructing dependency relationships, and enabling precise vulnerability impact assessment. It also demonstrates its superiority over existing state-of-the-art tools through a set of experiments in terms of runtime package recovery,  dependency-graph completeness, and vulnerability correlation. 

\subsection{Experimental Setup}
\label{s6_1}
We implemented \textsc{MEM-SBOM} as a suite of Volatility 3 (v2.26.2) plugins running on Ubuntu 20.04.6 LTS with 8 GB of RAM and Python 3.8.10. Application processes were identified using Volatility’s \texttt{pslist} and \texttt{pstree} plugins. The reachability analysis plugin takes as input the application's process ID,  main package (source), vulnerable package (target), vulnerable function, and the dependency graph, returning all runtime paths leading to the vulnerable code as output. The evaluated applications use Python versions ranging from 3.6 to 3.12. While the memory architecture remains stable across these versions, performance-oriented refinements introduce variations in internal offsets (e.g., the addresses of \texttt{\_PyRuntime} and \texttt{arenas}).  \textsc{MEM-SBOM} considers these version-specific offsets to maintain compatibility across all supported Python versions.

\subsubsection{Application Deployment}
We installed each of the applications using \texttt{pip install} into an isolated Python virtual environment created on Ubuntu 20.04 LTS virtual machines with  4 vCPUs and 4 GB RAM. 
This isolation ensures that each application is evaluated with its own dependency set, preventing any cross-application interference and guaranteeing that all modules recovered by \textsc{MEM-SBOM} belong solely to the target application. Memory dumps were captured by snapshotting the virtual machine immediately after the application completed its initialization phase, allowing sufficient time for it to complete startup and reach a steady execution state in which all the packages used by the application were resident in memory.  Modules related to environment management and packaging utilities (e.g., \texttt{pip}, \texttt{pkg\_resources}, \texttt{setuptools}, \texttt{wheel}, \texttt{distlib}, and \texttt{virtualenv}) as well as application-specific configuration files were excluded from comparisons to ensure fair comparison between the \textsc{MEM-SBOM} and SBOM tools.

\subsubsection{Dataset}
We evaluated 51 open-source Python applications from the \textit{Awesome-Python} repository \citep{awesome_python}. Each selected application constitutes a standalone system with its own dependency tree, metadata configuration, and runtime context, thereby capturing the diversity of Python packaging and execution models observed in practice.  The dataset covers a wide range of 23 categories, including CLI tools (e.g., \textit{iRedis}, \textit{LiteCLI}),  web frameworks (e.g., \textit{Django}, \textit{Flask}, \textit{FastAPI}), machine learning platforms (e.g., \textit{MLflow}, \textit{BentoML}, \textit{Rasa}), task schedulers (e.g., \textit{Apache Airflow}, \textit{Celery}, \textit{Prefect}), and data visualization tools (e.g., \textit{Apache Superset}, \textit{Orange}). This diversity reflects the heterogeneity of real-world Python deployments and enables evaluating \textsc{MEM-SBOM} under realistic operational conditions.  The complete dataset is presented in Table \ref{apps} of Appendix \ref{appendixA}.

\textbf{Note.} The versions installed for each application and its dependencies are determined by the underlying operating system and environment configuration. Consequently, the versions resolved by \texttt{pip} may not correspond to the most recent releases available on PyPI. Our dataset, therefore, reflects realistic deployment environments rather than enforcing the latest releases.

\subsubsection{Ground-Truth Generation}
 To evaluate the correctness of \textsc{MEM-SBOM} in recovering both installed package metadata and packages actively used at runtime, we constructed two complementary forms of ground truth: a static ground truth and a runtime ground truth. The static ground truth captures the set of packages installed in each application environment. We obtained this information using \texttt{pip freeze}, which enumerates all installed distributions and their resolved versions. This dataset serves as the reference for validating \textsc{MEM-SBOM}’s ability to reconstruct package names and versions directly from memory without filesystem access.
 
The runtime ground truth reflects the set of modules loaded by the Python interpreter during application execution. It is derived from the interpreter registry (\texttt{sys.modules}), which provides the authoritative record of legitimately imported modules in benign, non-adversarial deployments (see Section~\ref{s5_1_2}). Each application is instrumented using Python’s \texttt{sitecustomize} module, a built-in startup mechanism that the interpreter automatically imports before any application-level code \citep{python_site_module}, ensuring that our monitoring logic executes at interpreter initialization. The instrumentation captures the complete contents of \texttt{sys.modules} after the application reaches a steady execution state. It also reinitializes automatically in child processes spawned at runtime, providing complete coverage for multi-process applications.
When triggered, each process serializes its modules into a process-specific JSON file. Memory snapshots were acquired immediately thereafter, ensuring temporal alignment between the live module state and its in-memory representation. The same processing steps described in Sections \ref{s5_2_1} and \ref{s5_2_2} (i.e., module filtering and module grouping) are applied to the runtime ground-truth data to isolate the application modules.
This dual-ground-truth strategy provides complementary visibility into installed distributions and actively used packages, enabling accurate assessment of \textsc{MEM-SBOM}’s module-extraction accuracy. After validating \textsc{MEM-SBOM} against both views, we used it as the authoritative runtime baseline to evaluate the SBOM tools, while the static ground truth continues to serve as the reference for installed-package accuracy comparisons.

\subsubsection{SBOM Tools}
\begin{sloppypar}
We evaluated widely used SBOM tools: \textit{Syft} (v1.27.1) \citep{syft}, \textit{Trivy} (v0.65.0) \citep{trivy}, \textit{CDXGen} (v11.4.4) \citep{cdxgen}, \textit{CycloneDX-Python} (v7.1.0) \citep{cyclonedx_python}, \textit{SBOM4Python} (v0.12.4) \citep{sbom4python}, \textit{ORT} (v68.2.0) \citep{ort}, \textit{Jake} (v3.0.14) \citep{jake},  and \textit{PIP-SBOM} \citep{benedetti2024impact}.
For metadata-based evaluation, all eight tools were assessed. For deployment-based evaluation, we restricted the comparison to \textit{Syft}, \textit{CDXGen}, \textit{CycloneDX-Python}, and \textit{PIP-SBOM},  as these tools can analyze an existing installed environment or construct a synthetic one for SBOM generation. 
\textit{Syft} and \textit{CycloneDX-Python} directly inspect the evaluated environment by enumerating installed distributions from the \texttt{site-packages} directory. \textit{CDXGen}, in contrast, simulates an installation environment by creating a temporary virtual environment, resolving and installing all dependencies declared in the application’s manifests, and enumerating the resulting \texttt{site-packages} directory. \textit{PIP-SBOM}  does not inspect installed environments;  instead, it downloads the application’s distribution into a temporary directory, extracts its metadata, and resolves dependencies using PIP’s native logic. As a result, it reports the versions PIP would install, not the exact versions present in the evaluated environment.
\end{sloppypar}
\subsection{Evaluation Research Questions}
\label{s6_2}
Our evaluation addresses the following research questions:

\textbf{RQ1:} How accurately can \textsc{MEM-SBOM} recover the runtime state of Python applications, including installed, imported, and dynamically loaded packages?

\textbf{RQ2:} Can \textsc{MEM-SBOM} determine the actual reachability of known vulnerabilities within running applications?

\textbf{RQ3:} How do existing SBOM generation tools perform across diverse metadata configurations and real deployment environments, and how do their SBOMs impact the vulnerability detection accuracy compared to \textsc{MEM-SBOM}?

\subsubsection{Evaluation Metrics}
We evaluate \textsc{MEM-SBOM} and existing SBOM tools using four complementary categories of metrics that capture the package identification,  dependency graph construction, vulnerability detection accuracy, as well as vulnerability impact and reachability. All metrics described in this section are computed per application, and their mean values are calculated across the full evaluation dataset to summarize the aggregate performance of the tools and \textsc{MEM-SBOM}.

\paragraph{\textbf{Package Identification.}}
Let $P_S$, $P_T$, and  $P_M$ denote the sets of packages obtained from the static ground truth (\texttt{pip freeze}), the evaluated SBOM tool, and \textsc{MEM-SBOM}, respectively. 
We compute:
{\small
\begin{flalign*}
C_P &= \frac{|P_T \cap P_M|}{|P_M|} \times 100\%, \quad
M_P = \frac{|(P_S \cap P_M) \setminus P_T|}{|P_M|} \times 100\%, && \\[3pt]
R_P &= \frac{|P_M \setminus (P_S \cup P_T)|}{|P_M|} \times 100\% &&
\end{flalign*}}

\noindent
$C_P$ (\textit{Coverage}) measures the proportion of runtime packages correctly detected by the tool, 
$M_P$ (\textit{Missing}) quantifies packages that exist in both static and runtime environments but are missing from the tool output, and $R_P$ (\textit{Runtime-Only}) captures runtime-only dependencies invisible to static analysis. 

\paragraph{\textbf{Version accuracy.}} 
For each package present in all of the static ground truth ($P_S$),  the tool output ($P_T$), and the runtime set recovered by \textsc{MEM-SBOM} ($P_M$), we evaluate whether the version reported by the tool ($v_T(p)$) or by \textsc{MEM-SBOM} ($v_M(p)$) matches the static ground-truth version ($v_S(p)$). Version accuracy is the proportion of such matched versions and is defined as follows:

{\footnotesize
\begin{flalign*}
V_{acc} &=
\frac{|\{\,p \in (P_S \cap P_T \cap P_M) \mid v_X(p) = v_S(p)\,\}|}
     {|P_S \cap P_T \cap P_M|}
\times 100\%
\quad X \in \{T, M\} &&
\end{flalign*}
}

\paragraph{\textbf{Dependency Graph Construction.}}
For each package $p$, let $D_T(p)$ and $D_M(p)$ denote its direct dependencies in the graphs generated by the evaluated SBOM tool and \textsc{MEM-SBOM}, respectively, and let  $E_T$ and $E_M$ be the corresponding sets of directed edges. We define the common-edge domain ($E_{comm}$) as:

\begin{flalign*}
E_{\mathrm{comm}} \;=\; \{\, (u\!\to\!v) \in E_M \;:\; u,v \in V_T \cap V_M \,\}
\end{flalign*}

where $V_T$ and $V_M$ are the node sets of the tool and \textsc{MEM-SBOM} graphs, respectively. 
Using these definitions, we evaluate structural accuracy through three metrics defined as follows:

{\footnotesize
\begin{flalign*}
D_{acc} &= \frac{\sum_{p\in P_M} |D_T(p) \cap D_M(p)|}{\sum_{p\in P_M} |D_M(p)|} \times 100\%, \quad
&\hspace{-0.5cm} 
E_{acc} = \frac{|E_T \cap E_M|}{|E_{\text{comm}}|} \times 100\%, \\[4pt]
&\hspace{-0.5cm} 
G_{complete} = \frac{|E_T \cap E_M|}{|E_M|} \times 100\%
\end{flalign*}
}

\noindent
$D_{acc}$ (\textit{Dependency Accuracy}) measures how precisely a tool recovers direct runtime dependencies for packages observed in memory. 
$E_{acc}$ (\textit{Edge Accuracy}) measures how accurately the tool identifies runtime dependency edges for the packages that both the tool and  \textsc{MEM-SBOM} have in common ($E_{comm}$), while $G_{complete}$ (\textit{Graph Completeness}) evaluates global coverage by
comparing all runtime edges in memory with those reconstructed by the tool,
regardless of node presence.

\paragraph{\textbf{Vulnerability Detection.}}  
Let $V_S$, $V_T$, and $V_M$ denote the sets of vulnerabilities reported by  \textit{Grype} when analyzing SBOMs generated using \texttt{pip-audit}  (static ground truth), the evaluated tool, and \textsc{MEM-SBOM}, respectively. The static vulnerability baseline is obtained by executing \texttt{pip-audit} on the packages reported by  \texttt{pip freeze}.

{\footnotesize
\begin{flalign*}
C_V = \frac{|V_T \cap V_M|}{|V_M|} \times 100\%, \quad
P_V = \frac{|V_T \cap V_M|}{|V_T|} \times 100\%, \quad
\end{flalign*}
}
{\footnotesize
\begin{flalign*}
F_V = 2 \cdot \frac{C_V \cdot P_V}{C_V+ P_V},
\quad  M_V = \frac{|(V_S \cap V_M) \setminus V_T|}{|V_M|} \times 100\%,
\end{flalign*}
}
{\footnotesize
\begin{flalign*}
R_V = \frac{|V_M \setminus (V_S  \cup V_T)|}{|V_M|} \times 100\%.
\end{flalign*}
}

\noindent

\noindent $C_V$ (Coverage) measures the proportion of runtime vulnerabilities (i.e., those identified by \textsc{MEM-SBOM}) that are correctly identified by the tool. $P_V$ (\textit{Precision}) measures the proportion of tool-reported vulnerabilities that are present at runtime. 
$F_V$ (F1 Score) provides the harmonic mean of precision and recall, capturing the overall balance between detection completeness and correctness.
$M_V$ (\textit{Missing}) measures vulnerabilities that appear in both the static ground truth and memory but are missing from the tool’s output.
Finally, $R_V$ (\textit{Runtime-Only}) represents the fraction of vulnerabilities associated with runtime-only packages.

\paragraph{\textbf{Vulnerability Impact and Reachability.}}
To measure how a vulnerable dependency propagates through the application’s runtime graph, we employ three metrics.
\textbf{DEC (Direct Exposure Count)} measures the number of packages that directly depend on the vulnerable package, indicating the extent of immediate exposure. 
\textbf{RP (Reachability Percentage)} quantifies the fraction of all packages in the runtime graph that have a direct or transitive path to the vulnerable package, capturing the scope of potential impact.
\textbf{MDD (Minimum Dependency Distance)} reports the shortest path from the application's main package (source) to the vulnerable package (target), representing the depth of remediation required.

\subsection{Evaluation Results}
\label{s6_3}
This section presents and discusses the experimental results that answer the evaluation research questions.  

\subsubsection{RQ1: Package Recovery}
Table \ref{mem_sbom_1} summarizes the evaluation results of \textsc{MEM-SBOM}’s extraction accuracy using 23 representative applications, each chosen to illustrate a distinct category from the complete set of 51 applications. The results show the exact matching between the package metadata recovered from memory by inspecting the \texttt{\_path\_cache} structure and the static ground truth obtained from \texttt{pip freeze}. This validates the ability of \textsc{MEM-SBOM} to accurately reconstruct the name and version of the packages installed in the deployment environment without relying on filesystem artifacts. Similarly, the comparison between the runtime ground truth and the modules recovered from \texttt{sys.modules} yields a perfect match, demonstrating the accuracy of \textsc{MEM-SBOM} in capturing the subset of installed packages that the application actually imports during execution. As anticipated, not all installed packages are imported at runtime. This discrepancy arises from optional and conditional dependencies, platform-specific packages, and build-time utilities that are never used during execution. 
 For instance, in system-wide installations, the Python environment includes numerous OS-bundled packages that are present on the filesystem but never imported by the application.

\begin{sloppypar}
The results also illustrate a set of dynamic package modules that appear only at runtime. Such modules typically originate from packages dynamically loaded via \texttt{importlib.import\_module()} or \texttt{\_\_import\_\_()}, side-loaded packages bundled directly within the application directory, or plugin frameworks that introduce additional modules during execution. \textit{Glances} shows the largest runtime expansion (31 modules) due to its plugin architecture, which dynamically imports monitoring extensions from \texttt{glances/plugins/*.py} as independent top-level modules (e.g., \texttt{glances\_cpu}, \texttt{glances\_network}) via \texttt{importlib}, rather than hierarchical modules.  Meanwhile, \textit{Errbot}, \textit{Apache Superset}, and \textit{Modoboa} inject deployment-specific configurations through direct imports; \textit{Datasette},  \textit{Scapy}, and \textit{Mayan-EDMS}  expose backward-compatibility wrappers that create duplicated namespaces at runtime (e.g., \texttt{multipart}/\texttt{python\_multipart}, \texttt{attr}/\texttt{attrs}).  
\end{sloppypar}

\begin{table}[!htb]
\centering
\caption{Accuracy of \textsc{MEM-SBOM} in recovering installed, imported, and dynamically loaded packages.}
\label{mem_sbom_1}
\resizebox{\columnwidth}{!}{%
\begin{tabular}{c|cc|cc|c}
\toprule
& \multicolumn{2}{c|}{\textbf{Static Packages}} & \multicolumn{2}{c|}{\textbf{Imported Packages}} 
&\textbf{Dynamic} \\

\cmidrule(lr){2-3} \cmidrule(lr){4-5} 
\textbf{Application} & \textbf{Static} & \textbf{Memory}  & \textbf{Runtime}  & \textbf{Memory} & \textbf{Modules}  \\
& \textbf{Ground Truth} & \textbf{(\_path\_cache )} & \textbf{Ground Truth} & \textbf{(sys.modules)} & \\
\midrule
Ajenti-Panel & 50 & 50 &  40 &  40 &  0 \\
Apache Superset & 136 &  136 & 102  & 102  & 2  \\
APScheduler & 10 &10  &  9 & 9  & 0  \\
Beets &21 & 21 & 20  &  20 &  1  \\
Daphne &21 & 21 &  16 &  16 &  0 \\
Datasette & 27 & 27 & 21  & 21  & 1  \\
DevPi-Server & 66 & 66 &45   & 45  & 0   \\
Django & 5& 5 & 5  &  5 & 0 \\
Errbot&33 & 33 & 30  &  30 &  0 \\
FScociety & 14 & 14 &  12 & 12  & 0   \\
Glances  & 5 & 5 & 35  & 35  & 31   \\
iRedis &14 &  14 &  14 &  14 & 0  \\
JupyterLab &91 &91 & 69  & 69  &  0  \\
Lektor & 24 & 24 &  23 &  23 & 0 \\
Locust & 30 & 30 & 26  &  26  &  0 \\
Mayan-EDMS & 146 & 146 & 110  & 110  & 2    \\
MLflow &64 & 64 &  29 & 29  & 0 \\
Modoboa & 86 & 86 & 60  &  60 & 2  \\
RPyC & 4& 4  &  2 &  2 & 0  \\
RQ& 4 & 4 & 4  & 4  &  0 \\
Scapy & 1 & 1 &  1 &   1 &  0  \\
Scrapy &37 & 37  & 28  & 28  & 1   \\
Trytond&15 & 15 & 14  & 14  &  0  \\
\bottomrule
\end{tabular}%
}
\end{table} 

Table \ref{mem_sbom_2} highlights the robustness of \textsc{MEM-SBOM} across six heterogeneous installation environments for the \textit{Flask}  web framework, representing diverse packaging ecosystems. Despite substantial variation in dependency resolution and metadata formats (e.g., \texttt{.dist-info} and  \texttt{.egg-info}) across \texttt{pip}, \texttt{Conda}, \texttt{Poetry}, and system-level package managers, \textsc{MEM-SBOM} consistently enumerated the same eight imported packages of the application (i.e., \texttt{Flask}, \texttt{Itsdangerous}, \texttt{Jinja2}, \texttt{Blinker}, \texttt{click}, \texttt{Markupsafe}, \texttt{zipp}, and \texttt{Werkzeug}), demonstrating environment-independent extraction accuracy. However, the small differences in package counts reflect expected installation artifacts. Source-based installations introduce build tools (e.g.,  \texttt{wheel}) as a result of compiling the source archives using the package manager \texttt{pip}. \textit{Conda} inherits 78 user-scope packages unless \texttt{PYTHONNOUSERSITE=1} is set. \textit{Poetry} introduces a virtual infrastructure module (\texttt{\_virtualenv}) for environment isolation while \textit{APT}-based installations injects \texttt{apport\_python\_hook}, a system-level crash-reporting module. However, \textit{APT}-based installations show the highest heterogeneity, combining 76 user-level and 70 system-wide packages across \texttt{site-packages} and \texttt{dist-packages}. When Flask is installed system-wide, its version and dependency set depend on the operating system’s bundled Python packages. As a result, the application inherits legacy builds and distribution-specific variants (e.g., \texttt{Flask 1.1.1}) as distributed with Ubuntu 20.04, illustrating version divergence and mixed provenance that \textsc{MEM-SBOM} effectively exposes.

\begin{table}[!htb]
\centering
\caption{Accuracy of \textsc{MEM-SBOM} in recovering  the \textit{Flask}'s packages across heterogeneous installation environments.}

\label{mem_sbom_2}
\resizebox{\columnwidth}{!}{%
\begin{tabular}{c|c|cc|c|c}
\toprule
\textbf{Install} & \textbf{Environment} & \multicolumn{2}{c|}{\textbf{Static Packages}} & \textbf{Imported} & \textbf{Flask} \\
 \cmidrule(lr){3-4} 
\textbf{Method} & \textbf{Type} & \textbf{Env.} & \textbf{System}  & \textbf{Packages} &  \textbf{Version} \\
\midrule
pip (wheel, PyPI) & venv &12 & 0 &  8 &  3.0.3 \\
pip (source, PyPI) & venv & 13 & 0  & 8 &  3.0.3 \\
pip (GitHub) & venv & 12 & 0 & 8 & 3.0.3 \\
Conda (user inherit) & Conda + user scope & 0 & 78 &  8 &  3.0.3 \\
Poetry & poetry venv & 12 & 0 & 9 & 3.0.3 \\
System (APT) & System-wide & 0 & 146 &  9 & 1.1.1 \\
\bottomrule
\end{tabular}%
}
\end{table}

Beyond reconstructing an accurate runtime state from memory, \textsc{MEM-SBOM} identifies inconsistencies between the declared package version at installation and the version embedded in the loaded code. Table \ref{versions} highlights ten representative cases in our evaluation set where \textsc{MEM-SBOM} revealed such mismatches that metadata-based tools failed to detect.
These inconsistencies arise for several reasons. First, version attributes embedded in the source code may remain unchanged across releases, leading to incorrect version values in memory. Second, development builds may contain prerelease identifiers that are not reflected in the package manifests. Third, wrapper or alias artifacts may expose imported module names that diverge from the identifiers of the underlying distribution packages.

The Ajenti-Panel case exemplifies the third category, where \texttt{beautifulsoup4==4.13.4} provides the \texttt{bs4} import namespace, while a separate compatibility package \texttt{bs4==0.0.2} is installed in the same environment. This results in two distribution packages mapping to a single runtime namespace. Such mismatches can conceal vulnerable versions or enable dependency confusion, misleading tools that rely solely on static metadata. By validating version information directly from memory, \textsc{MEM-SBOM} can prove whether the installed packages align with their declared provenance, a critical capability for generating accurate SBOMs.

\begin{table}[!htb]
\centering
\caption{Version inconsistencies between package metadata and embedded code as detected by \textit{MEM-SBOM}.}
\label{versions}
\resizebox{\columnwidth}{!}{%
\begin{tabular}{llcc}
\toprule
\textbf{Application} & \textbf{Dependency} & \textbf{Installed Version} & \textbf{Recovered Version} \\
\midrule
Ajenti-Panel & beautifulsoup4 & 0.0.2 & 4.13.4 \\
DevPi-Server & strictyaml & 1.7.3 & 1.6.2 \\
Jet-Bridge & regex & 2024.11.6 & 2.5.148 \\
Lektor & lazy\_object\_proxy & 1.11.0 & 1.10.0 \\
Mayan-EDMS & regex & 2025.7.34 & 2.5.159 \\
Modoboa & oath & 1.4.4 & 1.4.0 \\
MyCLI & pymysql & 1.1.2 & 1.4.6 \\
Prefect & regex & 2025.9.1 & 2.5.161 \\
Rasa & terminaltables & 3.1.10 & 3.1.0 \\
Spyder & textdistance & 4.6.3 & 4.6.2 \\
\bottomrule
\end{tabular}
}
\end{table}

\begin{tcolorbox}[colback=gray!5!white, colframe=black!30, title=\textbf{Answer to RQ1}]
\textsc{MEM-SBOM} achieves 100\% coverage of the installed, imported, and dynamically loaded packages across heterogeneous deployment environments.
\end{tcolorbox}

\subsubsection{RQ2:  Vulnerability Reachability}
\begin{sloppypar}
Using \textit{Grype}, \textsc{MEM-SBOM} identifies all known vulnerabilities present in the evaluated applications and their dependent packages, including \textit{urllib3}, \textit{tornado}, and \textit{werkzeug}  as illustrated in Table \ref{vuls} of Appendix \ref{appendixA}. As a representative case study, we focused on the \textit{tornado} dependency, which is used by multiple applications and contains vulnerable routines located in the \texttt{tornado.httputil} module, namely \texttt{\_unquote\_cookie()}, \texttt{parse\_cookie()}, \texttt{parse\_body\_arguments()}, and \texttt{parse\_multipart\_form\_data()}. These functions are affected by \textit{CVE-2024-52804} (unsafe HTTP header parsing) and \textit{CVE-2025-47287} (unsafe multipart parsing). Despite the availability of patched releases, several applications continue to rely on outdated versions. We found that \textit{BentoML}, \textit{Flower}, \textit{JupyterLab}\textit{Jet-Bridge}, \textit{Spyder}, and \textit{Streamlit} embed \textit{tornado} 6.4.2, while  \textit{\textit{Jet-Bridge}} bundles \textit{tornado} 5.1.1.
\end{sloppypar}

To illustrate how this vulnerable dependency propagates through each application, Table \ref{tornado_1} reports the DEC (Direct Exposure Count), RP (Reachability Percentage), and MDD (Minimum Dependency Distance). All applications except \textit{BentoML} and \textit{Spyder} depend on \textit{tornado} directly (MDD=1). \textit{JupyterLab} exhibits the widest reach, with eight of its packages linked to \textit{tornado} (DEC=8, RP=14.3\%), whereas smaller frameworks such as \textit{Streamlit} and \textit{Jet-Bridge} show limited propagation (RP=2.9\% and 5.9\%, respectively). These variations illustrate how an application’s architecture determines the extent to which a vulnerable dependency propagates through its components. However, the presence of a vulnerable package does not necessarily imply that its vulnerable functionality is reachable from the application.
 
\begin{table}[!htb]
\centering
\caption{Evaluation of vulnerability propagation for the \textit{tornado} package across the evaluated applications.}
\label{tornado_1}
\resizebox{\columnwidth}{!}{%
\begin{tabular}{ccccccc}
\toprule
\textbf{Application} &  \textbf{Version} & \textbf{\#Pkgs} &\textbf{DEC} & \textbf{RP (\%)} &\textbf{MDD} & \textbf{Dep. Type} \\
\midrule
BentoML & 1.3.5 &  54 & 2 &   7.4 & \textbf{2} & Transitive \\
Flower & 2.0.1 & 19 & 1 &   5.3  & 1 & Direct \\
Jet-Bridge & 1.12.1 &  35 &  1&  2.9 & 1 & Direct \\
JupyterLab & 4.3.8 &  70 & \textbf{8} & \textbf{ 14.3} & 1 & Direct \\
Spyder & 6.0.8 & 105 & 3 &  6.7 & \textbf{2} & Transitive \\
Streamlit & 1.40.1 &  17 & 1 &   5.9 & 1 & Direct \\
\bottomrule
\end{tabular}
}
\end{table}

\begin{figure*}[!htb]
\centering
 \includegraphics[width=0.95\textwidth]{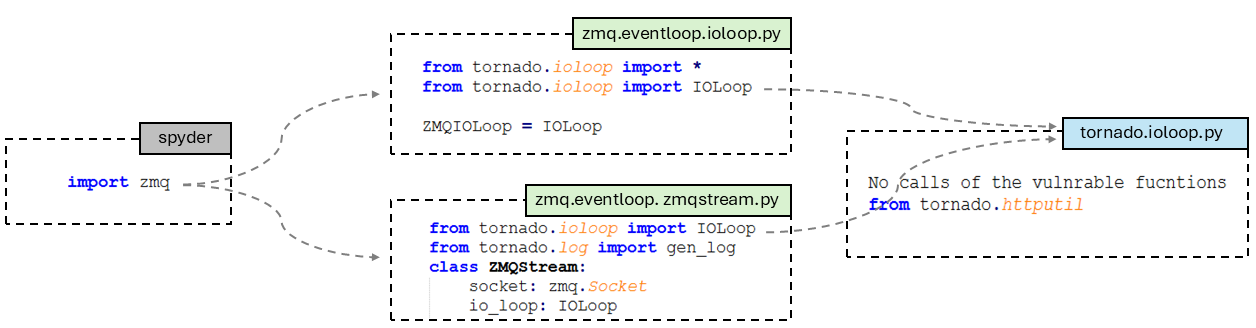}
\caption{Transitive dependency path from \textit{spyder} to the vulnerable \textit{tornado} package via \textit{zmq}, showing that no vulnerable functions are reached}
 \label{spyder}
\end{figure*}

\begin{figure*}[!htb]
\centering
 \includegraphics[width=0.95\textwidth]{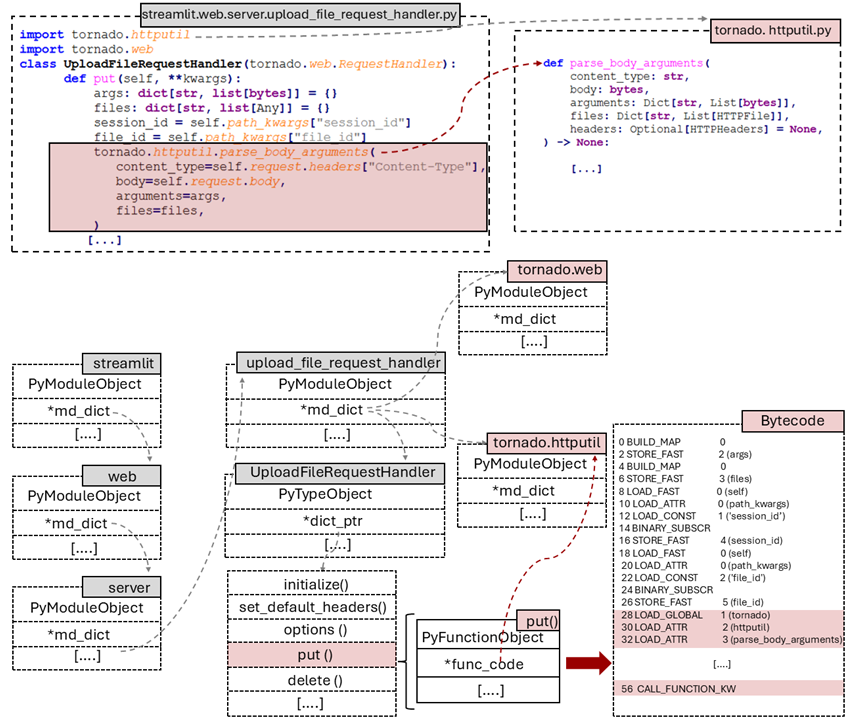}
 \caption{Direct call path from \textit{streamlit} to the vulnerable \texttt{tornado.httputil.parse\_body\_arguments()} function via the \texttt{put()} method, showing that the vulnerable function is reached}
 \label{streamlit2}
 \vspace{-0.1in}
\end{figure*}

Package-level scanning, therefore, overestimates risk. Figure \ref{spyder} illustrates that although \texttt{spyder} has a transitive dependency path to \texttt{tornado.ioloop} module via the \texttt{zmq} package, none of the vulnerable \texttt{httputil} functions are used. In contrast, Figure \ref{streamlit2} shows how \texttt{streamlit} reaches the vulnerable \texttt{parse\_body\_arguments()} function through its \texttt{UploadFileRequestHandler.put()} method, traversing the \texttt{web}, \texttt{server}, and \texttt{upload\_file\_request\_handler} modules. The bytecode of \texttt{put()} confirms a direct call to the vulnerable function.

\begin{sloppypar}
Table \ref{tornado_2} summarizes the function-level reachability results. While all six applications are flagged as vulnerable at the package level, the detailed inspection reveals that five of them never call the affected routines. For instance, \textit{Flower} imports several \textit{tornado} modules (e.g., \texttt{web}, \texttt{ioloop}, \texttt{httpserver}), yet none originate from \texttt{tornado.httputil}. \textit{JupyterLab} imports \texttt{httputil} but only uses benign routines such as \texttt{url\_concat()} and \texttt{HTTPHeaders.parse\_line()}, neither of which transitively invoke any vulnerable function. Only \textit{Streamlit} demonstrates a verifiable exploitation path, making it the only application where the vulnerability is reachable in memory. Thus, while package-based scanners flag 6/6 applications (100\%), function-level validation identifies only one true exposure (16.67\%), eliminating 83.3\% of false positives. This granularity shows that only \textit{Streamlit}’s upload handler requires remediation rather than upgrading \textit{tornado} across all dependent projects.
\end{sloppypar}

\begin{table*}[!htb]
\centering
\footnotesize
\caption{Function-level reachability analysis for the \textit{tornado} vulnerabilities. Only \textit{Streamlit} reaches a vulnerable function at runtime, while all other applications import only safe components.}
\label{tornado_2}
 \resizebox{\textwidth}{!}{%
\begin{tabular}{l|l|c|c|c}
\toprule
\textbf{Application} &
\textbf{Tornado Usage Summary} &
\textbf{Vulnerable Module Imported?} &
\textbf{Vulnerable Function Reached?} &
\textbf{Result} \\
\midrule
\textbf{BentoML} &
\texttt{ioloop}, \texttt{gen}, \texttt{concurrent} &
No & No &
\cellcolor{green!20}\textbf{Safe} \\
\textbf{Flower} &
\texttt{web}, \texttt{ioloop}, \texttt{httpserver}, \texttt{escape}, \texttt{log} &
No & No &
\cellcolor{green!20}\textbf{Safe} \\
\textbf{Jet-Bridge} &
\texttt{ioloop}, \texttt{gen}, \texttt{web}, \texttt{iostream} &
No & No &
\cellcolor{green!20}\textbf{Safe} \\
\textbf{JupyterLab} &
\texttt{httputil}, \texttt{url\_concat()}, \texttt{HTTPHeaders.parse\_line()} &
Yes & No &
\cellcolor{yellow!20}\textbf{Review} \\
\textbf{Spyder} &
\texttt{ioloop}, \texttt{httpclient} &
No & No &
\cellcolor{green!20}\textbf{Safe} \\
\textbf{Streamlit} &
\texttt{httputil.parse\_body\_arguments()} &
Yes & \cellcolor{red!20}\textbf{Yes (CVE-2025-47287)} &
\cellcolor{red!20}\textbf{Vulnerable} \\
\midrule
\multicolumn{5}{l}{\textbf{Summary:} Package-level scanning flag 6/6 (100\%), function-level scanning flags 1/6 (16.67\%).} \\
\bottomrule
\end{tabular}
}
\end{table*}

\begin{tcolorbox}[colback=gray!5!white, colframe=black!30, title=\textbf{Answer to RQ2}]
\textsc{MEM-SBOM} accurately identifies which applications reach vulnerable routines at runtime, revealing that only \textit{Streamlit} invokes the affected \textit{tornado} functions associated with \textit{CVE-2025-47287}, demonstrating precise vulnerability reachability at runtime.
\end{tcolorbox}

\subsubsection{RQ3: Comparison with SBOM Tools}
This section first analyzes the limitations of existing SBOM tools in parsing and supporting heterogeneous metadata formats across the evaluated applications. It then compares the runtime package coverage, dependency-graph accuracy, and vulnerability detection effectiveness of tools capable of analyzing deployed environments, using \textsc{MEM-SBOM} as the runtime ground truth.
\begin{sloppypar}
\noindent \paragraph{\textbf{Metadata-Based Analysis.}}
We collected applications from both \textit{PyPI} and \textit{GitHub} repositories. We observed that 30 \textit{GitHub} projects have a \texttt{requirements.txt} file, a format supported by all tools, which explains the higher success rates observed in the \textit{GitHub} dataset. The applications are grouped by metadata files into PEP 621-compliant projects (\texttt{pyproject.toml}), Poetry-managed projects (\texttt{pyproject.toml} + \texttt{poetry.lock}), and legacy or hybrid builds relying on \texttt{setup.py} and \texttt{setup.cfg}. The final group, \texttt{requirements.txt} (combined metadata), denotes projects that include a \texttt{requirements.txt} file alongside other metadata files. Tables \ref{tools_1} and \ref{tools_2} highlight systematic weaknesses in metadata parsing across all tools. For PEP 621 projects, six tools (\textit{Syft}, \textit{Trivy}, \textit{CDXGen}, \textit{CycloneDX-Python}, \textit{ORT}, and \textit{Jake}) fail completely in both datasets, while only \textit{PIP-SBOM} and \textit{SBOM4Python} succeed (100\%). In Poetry-managed projects, reliability depends almost entirely on the presence of a lock file. without \texttt{poetry.lock}, all tools except \textit{PIP-SBOM} fail to resolve dependencies; with it, most succeed except \textit{SBOM4Python}, which lacks native support for \texttt{[tool.poetry.dependencies]}. \textit{ORT} achieves partial success (66.7\%) owing to an outdated Poetry runtime that cannot parse the \texttt{priority} field introduced in version 1.2.
\end{sloppypar}
For legacy \texttt{setup.py}–based projects, success rates vary widely due to the imperative and non-standard nature of dependency declarations. \textit{Trivy}, \textit{CycloneDX-Python}, and \textit{Jake} provide no \texttt{setup.py} support (0\%). Among regex-based tools, \textit{Syft} captures only pinned dependencies (35.71\% PyPI, 20\% GitHub), while \textit{CDXGen} performs slightly better (42.86\% / 80\%) but fails on multiline or variable-defined lists and \texttt{extras\_require} fields. \textit{SBOM4Python} gives the highest coverage among pattern-matching tools (57.14 \% / 60\%) yet remains limited to single-line \texttt{install\_requires}, producing empty SBOMs when absent (e.g., \textit{Glances}). \textit{ORT} achieves comparable rates (42.86\% / 80\%) by executing \texttt{setup.py} and resolving dependencies through PyPI queries, a more flexible but nondeterministic strategy. Only \textit{SBOM4Python} and \textit{PIP-SBOM} correctly interpret auxiliary \texttt{setup.cfg} configurations, yielding complete outputs for hybrid declarative-imperative builds.

Although all evaluated tools support the \texttt{requirements.txt} file, they differ significantly in version handling and completeness. \textit{Syft} and \textit{Trivy} ignore unpinned dependencies (\texttt{==}), omitting over half of declared packages. \textit{CDXGen} retains all constraints, while \textit{SBOM4Python} and \textit{ORT} include unpinned entries but assign versions as \texttt{none}, relying respectively on regex parsing and the \texttt{pip-requirements-parser} library \citep{nexb_pip_requirements_parser}. \textit{Jake} and \textit{CycloneDX-Python} exhibit inconsistent behavior due to API divergence; older parser versions in \textit{Jake} omit unpinned entries, while newer releases used by \textit{CycloneDX-Python} include them unconditionally. None of the tools correctly interpret pip-specific directives such as \texttt{-r} (recursive inclusion) or \texttt{-e} (editable installations), resulting in systematically incomplete SBOMs.
Notably, \textsc{MEM-SBOM} is not affected by metadata configuration, producing precise runtime SBOMs that reflect actual execution.

These metadata-parsing limitations raise the question of whether SBOM tools can more accurately represent the deployed software state when analyzing installed environments rather than source metadata. To address this, we evaluate their performance in real runtime deployments. Notably, all subsequent runtime experiments are conducted on the PyPI-deployed versions of the applications. 
\begin{table*}[!htbp]
\centering
\caption{Success rates (\%) of SBOM tools across various metadata configurations for 51 Python applications (PyPI dataset).}
\label{tools_1}
\resizebox{\textwidth}{!}{%
\begin{tabular}{l|cccccccccc}
\toprule
\textbf{Metadata Configuration} & \textbf{\#Apps} & \textbf{Syft} & \textbf{Trivy} & \textbf{CDXGen} & \textbf{CycloneDX-Python} & \textbf{SBOM4Python} & \textbf{ORT} & \textbf{Jake} & \textbf{PIP-SBOM} & \textbf{MEM-SBOM} \\
\midrule
\makecell{pyproject.toml \\ (PEP 621 / setuptools)} & 8 & \makecell{0/8 \\(0\%)} & \makecell{0/8 \\ (0\%)} & \makecell{0/8 \\(0\%)} & \makecell{0/8 \\(0\%)} & \makecell{8/8 \\(100\%)} & \makecell{0/8 \\(0\%)} &\makecell{0/8 \\(0\%)}&\makecell{8/8 \\(100\%)} & \makecell{8/8 \\(100\%)} \\

\midrule
\makecell{pyproject.toml \\ (Poetry-managed)}  & 4 &  \makecell{0/4 \\(0\%)} & \makecell{0/4 \\(0\%)} & \makecell{0/4 \\(0\%)} & \makecell{0/4 \\(0\%)} & \makecell{0/4 \\(0\%)} & \makecell{0/4 \\(0\%)} & \makecell{0/4 \\(0\%)} & \makecell{4/4 \\(100\%)} & \makecell{4/4 \\(100\%)} \\

\midrule
\makecell{setup.py \\ $\pm$  setup.cfg} & 14 & \makecell{5/14\\ (35.71\%)} & \makecell{0/14 \\(0\%)} & \makecell{6/14 \\(42.86\%)} & \makecell{0/14\\ (0\%)} & \makecell{8/14 \\(57.14\%)} & \makecell{6/14 \\(42.86\%)} & \makecell{0/14 \\(0\%)} & \makecell{13/14\\ (92.86\%)} & \makecell{14/14\\ (100\%)} \\ 

\midrule
\makecell{pyproject.toml \\ + setup.cfg} & 4 & \makecell{0/4 \\(0\%)} & \makecell{0/4 \\(0\%)} & \makecell{0/4\\ (0\%)} & \makecell{0/4 \\(0\%)} & \makecell{4/4\\ (100\%)} & \makecell{0/4 \\(0\%)} & \makecell{0/4 \\(0\%)} & \makecell{4/4 \\ (100\%)} & \makecell{4/4 \\ (100\%)} \\

\midrule
\makecell{pyproject.toml \\ + setup.py $\pm$  setup.cfg}  & 7 & \makecell{1/7 \\(14.29\%)} & \makecell{0/7 \\ (0\%)} & \makecell{3/7 \\ (42.86\%)} & \makecell{0/7 \\ (0\%)} & \makecell{5/7 \\ (71.43\%)} & \makecell{2/7\\ (28.57\%)} & \makecell{0/7 \\ (0\%)} & \makecell{5/7 \\ (71.43\%)} & \makecell{7/7 \\ (100\%)} \\
\midrule
\makecell{requirements.txt \\$\pm$ (combined metadata)} & 14 & \makecell{10/14 \\ (71.43\%)} & \makecell{6/14 \\(42.86\%)} & \makecell{10/14 \\ (41.43\%)} & \makecell{14/14 \\ (100\%)} & \makecell{14/14 \\ (100\%)} & \makecell{7/14 \\ (50\%)} & \makecell{13/14 \\ (92.86\%)} & \makecell{14/14 \\ (100\%)} & \makecell{14/14 \\ (100\%)} \\
 \midrule
\end{tabular}%
}
\end{table*}

\begin{table*}[!htbp]
\centering
\caption{Success rates (\%) of SBOM tools across various metadata configurations for 51 Python applications (GitHub dataset).}
\label{tools_2}
\resizebox{\textwidth}{!}{%
\begin{tabular}{l|cccccccccc}
\toprule
\textbf{Metadata Configuration} & \textbf{\#Apps} & \textbf{Syft} & \textbf{Trivy} & \textbf{CDXGen} & \textbf{CycloneDX-Python} & \textbf{SBOM4Python} & \textbf{ORT} & \textbf{Jake} & \textbf{PIP-SBOM} & \textbf{MEM-SBOM} \\
\midrule
\makecell{pyproject.toml \\ (PEP 621 / setuptools)} &  4 & \makecell{0/4 \\(0\%)} & \makecell{0/4 \\(0\%)} & \makecell{0/4 \\(0\%)} & \makecell{0/4 \\ (0\%)} & \makecell{4/4 \\ (100\%)} & \makecell{0/4 \\ (0\%)} & \makecell{0/4 \\ (0\%)} & \makecell{4/4 \\ (100\%)} & \makecell{4/4 \\(100\%)} \\

\midrule
\makecell{pyproject.toml \\ (Poetry-managed)}  & 3 & \makecell{ 3/3\\ (100\%)}& \makecell{ 3/3 \\(100\%)} & \makecell{ 3/3 \\(100\%)} & \makecell{ 3/3 \\(100\%)} & \makecell{ 0/3 \\(0\%)} & \makecell{ 2/3 \\(66.67\%)} & \makecell{ 3/3 \\(100\%)} & \makecell{ 3/3 \\(100\%)} & \makecell{ 3/3\\ (100\%)} \\
\midrule
\makecell{setup.py \\ $\pm$  setup.cfg} &5 & \makecell{ 1/5 \\(20\%)} & \makecell{ 0/5 \\ (0\%)}& \makecell{ 4/5 \\ (80\%)} & \makecell{ 0/5 \\ (0\%)} & \makecell{3/5 \\ (60\%)} & \makecell{4/5 \\ (80\%)} & \makecell{  0/5\\(0\%)}& \makecell{ 5/5 \\(100\%)}& \makecell{ 5/5\\(100\%)} \\

\midrule
\makecell{pyproject.toml \\ + setup.cfg} &  2 & \makecell{ 0/2 \\(0\%)}& \makecell{ 0/2 \\ (0\%)} & \makecell{ 0/2 \\(0\%)} & \makecell{ 0/2 \\(0\%)} & \makecell{ 1/2 \\ (50\%)} & \makecell{ 0/2 \\ (0\%)} & \makecell{ 0/2 \\(0\%)} & \makecell{ 2/2 \\ (100\%)} & \makecell{ 2/2\\ (100\%)} \\ 

\midrule
\makecell{pyproject.toml \\ + setup.py $\pm$  setup.cfg}  & 7 & \makecell{ 0/7 \\ (0\%)} & \makecell{ 1/7\\ (14.29\%)} & \makecell{ 2/7 \\ (28.57\%)} & \makecell{ 0/7 \\ (0\%)} & \makecell{ 5/7 \\ (71.43\%)} & \makecell{ 3/7\\ (42.86\%)} & \makecell{ 0/7 \\ (0\%)} & \makecell{ 7/7 \\ (100\%)} & \makecell{ 7/7 \\ (100\%)} \\

\midrule
\makecell{requirements.txt \\$\pm$ (combined metadata)} & 30 & \makecell{ 23/30 \\(76.67\%)} & \makecell{ 17/30\\ (73.91\%)} & \makecell{ 27/30 \\(90\%)} & \makecell{ 29/30 \\(96.67\%)}& \makecell{ 30/30\\ (100\%)}& \makecell{ 18/30 \\(60\%)}& \makecell{ 30/30 \\(100\%)}& \makecell{ 25/30 \\(83.33\%)} & \makecell{ 30/30\\ (100\%)} \\ 

 \midrule
\end{tabular}%
}
\end{table*}

\noindent \paragraph{\textbf{Deployment-Based Analysis.}}
We evaluated how \textit{Syft}, \textit{CycloneDX-Python}, \textit{CDXGen}, and \textit{PIP-SBOM} tools perform in real deployment environments, where dependencies are installed, version-resolved, and dynamically loaded. This analysis verifies whether these tools can accurately represent the runtime software state and produce accurate dependency graphs, using \textit{MEM-SBOM} as the runtime ground truth. Given that these tools enumerate only direct dependencies, our comparison focuses on first-order dependency relationships, excluding transitive dependencies inferred through indirect imports.

As shown in Table~\ref{tools_3}, both \textit{Syft} and \textit{CycloneDX-Python} demonstrate stable behavior with high coverage (94.81\% and 90.9\%) and near-perfect version accuracy (>99.9\%) by directly inspecting installed distributions and parsing metadata from the \texttt{site-packages} directory of the active virtual environment. Consequently, they construct realistic dependency graphs that closely reflect the deployed state, achieving the highest dependency accuracy, edge accuracy, and overall graph completeness. 

In contrast, the predictive behavior of \textit{CDXGen} and \textit{PIP-SBOM} leads to the lowest performance across all metrics. Although \textit{CDXGen} infers dependencies by creating a temporary virtual environment and installing dependencies using the \texttt{--install-deps} flag, it still relies on static metadata analysis to construct its SBOM rather than observing actual runtime imports. \textit{PIP-SBOM} instead resolves constraints in metadata files by predicting what dependencies \emph{would} be installed rather than those actually present. This predictive strategy introduces false positives, omits conditional or optional dependencies, and frequently misreports versions when resolution assumptions diverge from the real deployment, resulting in substantially lower dependency accuracy ($D_{acc} < 50\%$) and incomplete graphs ($G_{complete} < 50\%$), consistent with their weaker runtime coverage. \textit{MEM-SBOM}, by directly enumerating the interpreter’s loaded modules, eliminates these limitations and achieves complete coverage (100\%) with 99.3\% version accuracy due to inconsistencies between the declared package version at installation and the version embedded in the loaded code (see Table \ref{versions}). However, all these tools are inherently static, missing packages that are dynamically loaded or generated at runtime.
 
\begin{table}[!htbp]
\centering
\caption{Performance of SBOM tools in deployment environments}
\label{tools_3}
\resizebox{\columnwidth}{!}{%
\begin{tabular}{l| c c c | c| c c c}
\toprule
\textbf{Tool} & $C_P$ & $M_P$ & $R_P$ & $V_{acc}$ & $D_{acc}$ & $E_{acc}$ & $G_{complete}$ \\
& (\%) & (\%) & (\%) & (\%) & (\%) & (\%) & (\%) \\
\midrule
Syft & 94.81 & 0.90 & 4.29 & 99.96 & 73.81 & 71.76 & 62.24 \\
CDXGen & 76.55 & 17.69 & 5.76 & 52.11 & 48.05 & 54.89 & 49.63 \\
CycloneDX-Python & 90.90 & 4.45 & 4.65 & 99.98 & 72.23 & 75.52 & 71.66 \\
PIP-SBOM & 73.42 & 21.96 & 4.61 & 65.22 & 27.90 & 34.15 & 20.09 \\
MEM-SBOM & 100.00 & 0.00 & 0.00 & 99.32 & 100.00 & 100.00 & 100.00 \\
\bottomrule
\end{tabular}%
}
\end{table}

Since \textit{Grype} relies on the generated SBOMs, any inaccuracies in these SBOMs directly affect the reported vulnerability set, as illustrated in Table~\ref{tools_4}. \textit{Syft} achieves full coverage ($C_V=100\%$) equivalent to \textit{MEM-SBOM}, while \textit{CycloneDX-Python} provides near-complete detection (96.3\%), missing only 3.7\% of runtime vulnerabilities. In contrast, \textit{CDXGen} and \textit{PIP-SBOM} exhibit coverage below 40\% and high omission ratios ($R_V>60\%$), consistent with their incomplete runtime visibility. Notably,  $M_V$ is zero for all tools, indicating that vulnerabilities reported by both \texttt{pip audit} and \textit{MEM-SBOM} are consistently detected; the missed vulnerabilities arise entirely from runtime-only packages. Finally, $P_V$ decreases when tools report vulnerabilities associated with packages listed in metadata but never imported at runtime, thereby introducing false positives and reducing the overall F1 score ($F_V$). By restricting analysis to modules observed in memory, \textit{MEM-SBOM} eliminates such false positives and achieves perfect precision, recall, and F1 ($P_V=R_V=F_V=100\%$).

\begin{table}[!htb]
\centering
\tiny
\caption{The impact of SBOM tools on vulnerability detection.}
\label{tools_4}
\resizebox{\columnwidth}{!}{%
\begin{tabular}{lccccc}
\toprule
\textbf{Tool} & \textbf{Coverage} & \textbf{Missing} & \textbf{Runtime\_Only} & \textbf{Precision}& \textbf{F1 Score}\\
              & \textbf{(\%)} & \textbf{(\%)} & \textbf{(\%)} & \textbf{(\%)} & \textbf{(\%)} \\
\midrule
Syft          & 100.00 & 0.00 & 0.00 & 75.07 &  80.00 \\
CDXGen        &27.48 & 0.00 &72.52 & 23.58 &  23.70 \\
CycloneDX-Python     &96.30 & 0.00 & 3.70 & 88.49 & 91.35 \\
PIP-SBOM     & 37.37 & 0.00 & 62.63 & 33.80 &  34.67 \\
MEM-SBOM      & 100.00 & 0.00 & 0.00 &  100.00 & 100.00 \\
\bottomrule
\end{tabular}
}
\end{table}

\begin{tcolorbox}[colback=gray!5!white, colframe=black!30, title=\textbf{Answer to RQ3}]
Across heterogeneous metadata configurations and deployed environments, existing SBOM tools remain constrained by the metadata they parse, providing only partial visibility into the application runtime state.  As a result, they generate incomplete SBOMs and dependency graphs, which in turn impact the accuracy of vulnerability detection.
\end{tcolorbox}
\section{Limitations and Future Work}
\label{s7}
This work focuses primarily on the Python runtime, excluding native C/C++ extensions that operate outside Python’s managed memory. Such extensions introduce additional attack surfaces through language boundary transitions and type conversions \citep{scholtes2025charon}. Extending \textit{MEM-SBOM} to support polyglot correlation between Python objects and native components would enhance cross-language visibility. \textit{MEM-SBOM} currently analyzes only components resident in process memory at the time of acquisition. Therefore,  attacks that occur solely during installation remain unobserved. Integrating installation-phase monitoring or provenance tracing could capture such transient artifacts produced during setup routines and extend visibility beyond the runtime state. Vulnerability assessment depends on Grype’s aggregation of public databases (e.g., NVD, OSV, and GitHub Security Advisories), which limits detection to known CVEs.  Future work may incorporate semantic pattern analysis and anomaly detection over memory-derived artifacts to identify previously unseen vulnerabilities. Finally, this work assumes that the Python runtime environment itself is not malicious or compromised. Prior work \citep{park2018bytecode} showed that memory-safety flaws in interpreters can corrupt bytecode while preserving valid metadata. Deep inspection of opcode sequences and Python-to-native wrapper transitions could further expose such tampering and injected behaviors.
\section{Conclusion}
\label{s8}
In this paper, we introduced \textit{MEM-SBOM}, the first memory forensics framework that generates accurate SBOMs directly from the runtime state of Python applications. By traversing the interpreter registries, thread states, garbage collector, arenas, and heap regions, the framework recovers all resident modules, even under evasive conditions. The extracted modules are then filtered and grouped by their parent packages to isolate the application packages, whose versions are determined through a multi-source resolution process. The resulting packages are then analyzed at the object level to construct dependency graphs and perform fine-grained reachability analysis.
Evaluation across 51 real-world Python applications demonstrated that \textit{MEM-SBOM} achieves 100\% extraction accuracy and detects ten inconsistency cases between the declared package version at installation and the version embedded in the loaded code. In a detailed analysis of six \textit{tornado}-dependent applications, \textit{MEM-SBOM} identifies \textit{Streamlit} as the only application that executes the vulnerable routines at runtime, eliminating 83.3\% of package-level false positives. It also recovers all runtime-only dependencies, 4--6\% of which are consistently missed by existing tools, resulting in more complete dependency graphs and more accurate vulnerability correlation.
These results show that memory-based SBOMs provide the runtime visibility essential for software supply chain security and for effective incident response, particularly in dynamic language ecosystems where metadata and filesystem artifacts fail to reveal the actual runtime state.

\section{Code availability}
\textit{MEM-SBOM} is implemented as a suite of open-source Volatility 3 plugins, publicly available at \url{https://github.com/HalaAli198/MEM-SBOM}.

\bibliographystyle{cas-model2-names}
\bibliography{main}
\appendix
\section{Appendix}
\label{appendixA}
\begin{table*}[!htbp]
\footnotesize
\centering
\caption{Evaluation dataset of 51 Python applications spanning 23 categories.}
\label{apps}
\scriptsize
\begin{tabular}{l|c}
\toprule
\textbf{Application Category} & \textbf{Application Name (Version)} \\ 
\midrule
Web Frameworks & Django (v4.2.23), Flask (v3.0.3), FastAPI (v0.116.1) \\
\midrule
Web-based Environments & Jupyter Notebook (v7.4.5), JupyterLab (v4.3.8) \\
\midrule
Task Scheduling Systems & Apache Airflow (v3.0.0), Celery (v5.5.3), APScheduler (v4.0.0a5), Prefect (v3.4.17) \\
\midrule
Machine Learning Platforms & MLflow (v2.17.2), BentoML (v1.3.5), Seldon Core (v1.18.2), Rasa (v3.6.21), Gradio (v4.44.1) \\
\midrule
Data Visualization & Apache  Superset (v2.1.3), Orange (v3.38.1) \\
\midrule
ASGI Servers & Daphne (v4.1.2), Uvicorn (v0.33.0), Hypercorn (v0.17.3) \\
\midrule
Admin Panels & Ajenti-Panel (v2.2.11), Flower (v2.0.1), Jet-Bridge (v1.12.1), Wooey (v0.13.3), Streamlit (v1.40.1) \\
\midrule
CLI  Tools & iRedis (v1.15.2), LiteCLI (v1.13.2), MyCLI (v1.30.0) \\
\midrule
Static Site Generators & Lektor (v3.3.12), MkDocs (v1.6.1), Pelican (v4.10.2), Nikola (v8.3.3) \\
\midrule
Email Servers & Modoboa (v2.3.6), Salmon (v3.3.0) \\
\midrule
Development Tools & Supervisor (v4.3.0), DevPi-Server (v6.15.0), Spyder (v6.0.8) \\
\midrule
Database Tools & SQLite-Web (v0.6.4), Datasette (v0.64.3) \\
\midrule
RPC Servers & RPyC (v6.0.2), Zerorpc (v0.6.3) \\
\midrule
System Monitoring & Glances (v3.4.0.3) \\
\midrule
Penetration Testing & FSociety (v3.2.9) \\
\midrule
Network Tools & Mininet (v2.3.0.dev6), Scapy (v2.6.1) \\
\midrule
Web Scraping & Scrapy (v2.11.2) \\
\midrule
Audio Management & Beets (v2.2.0) \\
\midrule
Chatbots & Errbot (v6.2.0) \\
\midrule
Business Software & Trytond (v6.6.1) \\
\midrule
Document Management & Mayan-EDMS (v4.9.2) \\
\midrule
Task Queues & RQ (v2.3.3) \\
\midrule
Testing Tools & Locust (v2.25.0) \\
\bottomrule
\end{tabular}
\end{table*}
\begin{table*}[!htbp]
\centering
\caption{Detected vulnerable packages across the evaluated Python applications.}
\label{vuls}
\scriptsize
\begin{tabular}{llccccccc}
\toprule
\textbf{Project} &\textbf{Version} & \textbf{Affected Package} & \textbf{Package Version} & \textbf{Critical} & \textbf{High} & \textbf{Medium} & \textbf{Low}  \\
\midrule
Ajenti-Panel & 2.2.11 & urllib3 & 2.2.3 & -- & -- &  2 & -- \\
\midrule

BentoML & 1.3.5 
     & bentoml & 1.3.5 & 3 & 2 & 1 & -- \\
    & & tornado & 6.4.2 & -- & 1 & -- & -- \\
    & & starlette & 0.44.0 & -- & -- & 1 & --  \\

\midrule
DevPi-Server & 6.15.0
    & waitress & 3.0.0 & 1 & 1 & -- & -- \\
    && urllib3 & 2.2.3 & -- & -- & 2 & --  \\

\midrule
Django & 4.2.23 & django & 4.2.23 & -- & 1 & -- & --  \\

\midrule
Errbot & 6.2.0 
    & cryptography & 41.0.7 & -- & 2 & 2 & --  \\
    & & jinja2 & 3.1.2 & -- & -- & 5 & --  \\
   &  & requests & 2.31.0 & -- & -- & 2 & --  \\

\midrule
FastAPI & 0.116.1 
    & starlette & 0.44.0 & -- & -- & 1 & --  \\
   &  & urllib3 & 2.2.3 & -- & -- & 2 & -- \\

\midrule
Flower & 2.0.1 & tornado & 6.4.2 & -- & 1 & -- & --  \\

\midrule
Fsociety & 3.2.9 & urllib3 & 2.2.3 & -- & -- & 2 & --  \\

\midrule
Gradio & 4.44.1 
    & gradio & 4.44.1 & 1 & 7 & 7 & --  \\
    &  & starlette & 0.44.0 & -- & -- & 1 & --  \\
    &  & urllib3 & 2.2.3 & -- & -- & 2 & -- \\
\midrule

Hypercorn &  0.17.3 & h2 & 4.1.0 & -- & -- & 1 & -- \\
\midrule
Jet-Bridge & 1.12.1 
    & tornado & 5.1.1 & -- & 2 & 4 & -- \\
  &  & pymongo & 4.1.1 & -- & -- & 1 & -- \\
   & & urllib3 & 2.2.3 & -- & -- & 2 & --  \\

\midrule
JupyterLab &4.3.8
    & tornado & 6.4.2 & -- & 1 & -- & --  \\
   & & urllib3 & 2.2.3 & -- & -- & 2 & --  \\
\midrule
Jupyter Notebook & 7.4.5 & jupyterlab & 4.4.5 & -- & -- & -- & 1 \\

\midrule
Lektor &  3.3.12 
    & werkzeug & 2.3.8 & -- & 1 & 2 & --  \\
&    & urllib3 & 2.2.3 & -- & -- & 2 & --  \\

\midrule
Locust & 2.25.0 
    & flask\_cors & 5.0.0 & -- & -- & 3 & --  \\
   &  & urllib3 & 2.2.3 & -- & -- & 2 & --  \\

\midrule
Mayan-EDMS & 4.9.2
    & django & 4.2.23 & -- & 1 & -- & --  \\
   & & pypdf & 5.1.0 & -- & -- & 1 & --  \\
 &  & urllib3 & 1.26.20 & -- & -- & 1 & --  \\

\midrule
MLflow & 2.17.2
    & mlflow & 2.17.2 & -- & -- & 3 & --  \\
  &   & urllib3 & 2.2.3 & -- & -- & 2 & --  \\

\midrule
MyCLI & 1.30.0
    & sqlparse & 0.4.4 & -- & 1 & -- & -- \\
   &  & paramiko & 2.11.0 & -- & -- & 1 & -- \\
\midrule
Nikola & 8.3.3 & urllib3 & 2.2.3 & -- & -- & 2 & --\\

\midrule
Rasa & 3.6.21
    & keras & 2.12.0 & 1 & 1 & 1 & -- \\
  & & tensorflow & 2.12.0 & -- & 1 & -- & --  \\
  &  & future & 1.0.0 & -- & 1 & -- & -- \\
  &  & aiohttp & 3.9.5 & -- & -- & 1 & --  \\
  &  & pymongo & 4.3.3 & -- & -- & 1 & -- \\
  &  & urllib3 & 1.26.20 & -- & -- & 1  & -- \\

\midrule
Scrapy & 2.11.2 & urllib3 & 2.2.3 & -- & -- & 2 & --  \\
\midrule
Seldon Core & 1.18.2
    & flask\_cors & 3.0.10 & -- & 1 & 4 & --  \\
  &  & gunicorn & 20.1.0 & -- & 2 & -- & -- \\
  &  & werkzeug & 2.2.3 & -- & 1 & 3 & -- \\

\midrule
Spyder & 6.0.8 
    & tornado & 6.4.2 & -- & 1 & -- & -- \\
 &    & urllib3 & 2.2.3 & -- & -- & 2 & --  \\
 &    & aiohttp & 3.10.11 & -- & -- & -- & 1  \\

\midrule
Streamlit &  1.40.1
    & tornado & 6.4.2 & -- & 1 & -- & -- \\
 &   & urllib3 & 2.2.3 & -- & -- & 2 & -- \\

\midrule
Superset & 2.1.3
    & pyarrow & 10.0.1 & 1 & -- & -- & -- \\
  &  & cryptography & 39.0.2 & -- & 2 & 3 & 3 \\
  &  & sqlparse & 0.4.4 & -- & 1 & -- & -- \\
  &  & werkzeug & 2.3.3 & -- & 1 & 3 & -- \\
  &  & flask\_caching & -- & -- & -- & 1 & -- \\
  &  & urllib3 & 2.2.3 & -- & -- & 2 & --  \\

\midrule
Zerorpc & 0.6.3 & future & 1.0.0 & -- & 1 & -- & --  \\
\bottomrule
\end{tabular}
\end{table*}

\end{document}